\documentclass[a4paper, 11pt]{article}
\usepackage{aapreprint}
\usepackage{amsmath}
\usepackage{amsthm}
\usepackage{xparse}

\usepackage[english]{babel}
\usepackage[utf8]{inputenc}
\usepackage[colorinlistoftodos]{todonotes}
\usepackage{hyperref}
\usepackage{graphicx,wrapfig,float,slashed}
\usepackage{amsmath,amssymb,dsfont,epsfig,graphicx,xcolor}
\usepackage{physics}
\usepackage{mathtools} 
\usepackage{multirow}
\usepackage{blkarray}
\usepackage{epstopdf}
\usepackage{ragged2e}
\usepackage[utf8]{inputenc} 
\usepackage{cancel}
\usepackage[toc]{appendix}
\usepackage[normalem]{ulem}
\usepackage{enumitem}
\usepackage[font=small,skip=1pt]{caption}
\allowdisplaybreaks
\usepackage{pifont}
\usepackage{subcaption}
\usepackage{footnote}

\usepackage[compat=1.0.0]{tikz-feynman}
\usepackage{cleveref}

\newcommand{\nc}{\newcommand}
\nc{\ut}[2][]{\ensuremath{~\textrm{#2}^{#1}}}
\nc{\non}{\nonumber}
\nc{\hc}{\hbox {h.c.}}
\nc{\noi}{\noindent}
\nc{\barx}{\bar{x}}
\nc{\pbarn}{\;\hbox {pb}}
\nc{\fbarn}{\;\hbox {fb}}
\nc{\val}[3][]{\ensuremath{\qty(\frac{#2}{#3})^{#1}}}

\nc{\hsp}{\hspace{0.5cm}}
\nc{\lsp}{\hspace{1cm}}
\nc{\Lsp}{\hspace{2cm}}
\nc{\LLsp}{\lsp\lsp}
\nc{\lra}{\longrightarrow}
\nc{\p}{\prime}
\nc{\sgn}{\text{sgn}}
\nc{\ph}{\varphi}
\nc{\vmol}{v_{\text{M\o l}}}
\nc{\sun}{\odot}
\nc{\msun}{\ensuremath{\mathrm{M}_{\sun}}}

\nc{\beq}{\begin{equation}}
\nc{\eeq}{\end{equation}}
\nc{\bea}{\begin{eqnarray}}  
\nc{\eea}{\end{eqnarray}}
\nc{\baa}{\begin{array}}     \nc{\eaa}{\end{array}}
\nc{\bit}{\begin{itemize}}   \nc{\eit}{\end{itemize}}
\nc{\ben}{\begin{enumerate}} \nc{\een}{\end{enumerate}}
\nc{\bce}{\begin{center}}    \nc{\ece}{\end{center}}
\nc{\bpm}{\begin{pmatrix}}   \nc{\epm}{\end{pmatrix}}
\nc{\bvt}{\begin{verbatim}}  
	\nc{\evt}{\end{verbatim}}

\title{Long-Lived-Particle Signals of a Composite Hidden Sector through the Neutrino Portal}

\author[1]{Aqeel Ahmed,}
\emailAdd{aqeel.ahmed@mpi-hd.mpg.de}
\author[2]{Zackaria Chacko,}
\emailAdd{zchacko@physics.umd.edu}
\author[3]{Niral Desai,}
\emailAdd{npd393@utexas.edu}
\author[2]{Sanket Doshi,}
\emailAdd{sdoshi@umd.edu}
\author[3]{\\ Can Kilic,}
\emailAdd{kilic@physics.utexas.edu}
\author[1]{Saereh Najjari,}
\emailAdd{saereh.najjari@mpi-hd.mpg.de}
\author[3]{Ram Purandhar Reddy Sudha}
\emailAdd{ramreddy@utexas.edu}
\affiliation[1]{Max-Planck-Institut für Kernphysik,\\ Saupfercheckweg 1, 69117 Heidelberg, Germany}
\affiliation[2]{Maryland Center for Fundamental Physics, Department of Physics,\\ University of Maryland, College Park, MD 20742-4111 USA}
\affiliation[3]{Theory Group, Weinberg Institute for Theoretical Physics,\\ University of Texas at Austin, Austin, TX 78712, USA}

\abstract{We explore the signals of a scenario in which the composite states of a strongly coupled hidden sector couple to the Standard Model through the neutrino portal, giving rise to the neutrino masses. We consider a framework in which the hidden sector is conformal in the ultraviolet and the compositeness scale lies below the weak scale. If the lightest composite state in the hidden sector is a scalar, its decay rate back to the Standard Model is suppressed by angular momentum considerations and can naturally be small, giving rise to long-lived particle signals. We determine the current constraints on this class of models and explore the reach of future collider and beam dump searches. We find that FASER, SHiP, and Belle II can potentially probe a significant part of the unexplored parameter space. 	
}

\begin{document}
\preprint{UT-WI-28-2025}

\maketitle
\section{Introduction}

The existence of a hidden sector composed entirely of particles that carry no charge under the Standard Model (SM) gauge groups is one of the most exciting possibilities for new physics. Such hidden sectors can resolve some of the most puzzling problems of the SM, including the little hierarchy problem~\cite{Chacko:2005pe,Barbieri:2005ri,Craig:2015pha,Cohen:2018mgv,Cheng:2018gvu}, the nature of dark matter~\cite{Silveira:1985rk,McDonald:1993ex,Burgess:2000yq,Dodelson:1993je,Pospelov:2007mp,Feng:2008mu}, the origin of neutrino masses~\cite{Minkowski:1977sc,Yanagida:1979,Gell-Mann:1979vob,Glashow:1979,Mohapatra:1979ia}, and the origin of the baryon asymmetry~\cite{Fukugita:1986hr,Luty:1992un,Farina:2016ndq,Feng:2020urb,Kilic:2021zqu}.
Since the experimental bounds on particles that are neutral under the SM are comparatively weak, the hidden sector states could have masses well below the weak scale. Consequently, the existence of such a hidden sector would have important implications for current and near-future experiments operating at these energies.

Hidden sectors interact with the SM through terms in the Lagrangian of the form $\mathcal{O}_{\rm SM}\mathcal{O}_{\rm HS}$, which couple gauge invariant SM operators ${\mathcal O}_{\rm SM}$ to operators $\mathcal{O}_{\rm HS}$ in the hidden sector. The SM operators ${\mathcal O}_{\rm SM}$ constitute ``portals'' to the hidden sector. The hidden sector may contain just a small number of weakly coupled states, but it could also contain a multitude of states that interact strongly with each other and exhibit rich dynamics.  

The experimental signals associated with a hidden sector depend on the portal through which it couples to the SM, and also on its particle content.
Portal operators of low scaling dimension are of particular interest since they can be probed with the greatest experimental sensitivity. The three lowest dimensional portal operators are the Higgs portal ${\mathcal O}_{\rm SM} \equiv H^{\dag}H$, the neutrino portal ${\mathcal O}_{\rm SM} \equiv \ell H$, and the hypercharge portal ${\mathcal O}_{\rm SM} \equiv B_{\mu\nu}$. For each of these portals, the experimental signals of the case when the hidden sector consists of just a few weakly interacting states have been carefully studied. These include searches for scalar particles that mix with the Higgs boson, fermions that couple through the neutrino portal, and dark photons that mix with the hypercharge gauge boson. However, the experimental signals of more general hidden sectors have received much less attention. 

In this paper, we explore the experimental signals arising from a strongly coupled hidden sector that couples to the SM through the neutrino portal and gives rise to the neutrino masses. We consider a framework in which the hidden sector is approximately conformal in the ultraviolet, but a deformation by a relevant operator causes it to confine in the infrared. The compositeness scale is taken to lie below the weak scale. The neutrino masses are generated through the inverse seesaw mechanism~\cite{Mohapatra:1986aw,Mohapatra:1986bd}, with the smallness of the neutrino masses arising from the scaling dimensions of operators in the conformal field theory (CFT). This framework for neutrino masses has been studied in Refs. \cite{Chacko:2020zze,Ahmed:2023vdb,Hong:2024zsn,Borrello:2025hal}, see also \cite{McDonald:2010jm}. In general, the experimental signals of such a scenario depend on the quantum numbers of the lightest states in the hidden sector. This is because, once produced, the heavier states in the hidden sector promptly cascade down to the lightest states that are kinematically forbidden from further decays into the hidden sector. These states must either be completely stable or decay into final states that contain SM particles. Therefore, the nature of the experimental signals depends sensitively on the quantum numbers of the lightest particles in the hidden sector and their decay modes.    

If the lightest state in the hidden sector is a composite fermion that inherits a neutrino portal interaction with the SM, its decays are identical to those of a conventional heavy neutral lepton (HNL). However, since the theory in the ultraviolet is strongly coupled, a single event at a collider or beam dump can now give rise to multiple composite fermions, and so the signals still differ from those of weakly coupled hidden sector models. This scenario was explored in Ref.~\cite{Chacko:2020zze}. The case in which the lightest particle in the hidden sector is stable and constitutes dark matter has been explored in Refs.~\cite{Ahmed:2023vdb,Hong:2024zsn}. Distinctive signals of the scenario in which the decays of the hidden sector are completely invisible were explored in Ref.~\cite{Borrello:2025hal}. The implications of couplings to a conformal hidden sector for neutrino self-interactions were considered in \cite{Foroughi-Abari:2025upe}. In this paper, we consider the scenario in which the  
the lightest composite state in the hidden sector is a scalar. In this framework, the decay rate of the scalar back to the SM is suppressed by angular momentum considerations and can naturally be small, giving rise to distinctive long-lived particle signals. We determine the current constraints on this class of models and explore the reach of future collider and beam dump searches. We find that FASER, SHiP, and Belle II can potentially probe a significant part of the unexplored parameter space.  

The outline of this paper is as follows. In the next section, we present the framework that underlies the class of models we are considering. In Section 3, we determine the current bounds on this scenario from colliders and beam dumps and explore the reach of future searches. We study the cosmological constraints on this class of models in Section 4 and the astrophysical limits in Section 5. We conclude in Section 6. 

\section{The Framework}

\label{sec: model}

In this section, we introduce a class of models that realize the scenario we are considering. We consider a strongly coupled hidden sector that couples to the SM through the neutrino portal. For concreteness, we take the strong dynamics to be that of a CFT deformed by a scalar operator $\mathcal{O}_S$ of dimension $\Delta_S$, where $\Delta_S < 4$. In the ultraviolet, the corresponding terms in the Lagrangian take the form,
 \begin{align}
 \mathcal{L}_{\rm UV} \supset \mathcal{L}_{\rm CFT} + \lambda_S \mathcal{O}_S \;.
 \label{LagCFT}
 \end{align}
When the deformation $\mathcal{O}_S$ gets large, it triggers breaking of the CFT at a scale which we denote by $f$. We focus on a scenario in which there are no pions or other light composite states, so the lightest particles in the hidden sector have masses of order the compositeness scale $\Lambda \equiv 4 \pi f$. We restrict our attention to the case when the compositeness scale lies below the weak scale, which gives rise to distinctive signals.  

It is conventional to normalize the operators in a CFT based on their two-point functions. For scalar operators such as $\mathcal{O}_S$, unitarity places a lower bound on their scaling dimensions, $\Delta_S \geq 1$. For the range of scaling dimensions $1 \leq \Delta_{S} < 2$, we can follow the conventions of unparticle physics~\cite{Georgi:2007ek,Georgi:2007si} to normalize the operator $O_S$,
 \begin{equation}
\int d^4x \; e^{ipx}
\langle 0|T\left[ \mathcal{O}_{S}^{\dagger}(x) \mathcal{O}_{S}(0)\right] |0 \rangle =
-\frac{A_{\Delta_{S}}}{2 i\, {\rm sin}
\left( \Delta_{S} \pi \right) }
\frac{1}
{\left(-p^2 - i \epsilon\right)^{2 - \Delta_{S}} }~,
\label{georgi-norm}
 \end{equation}
where the normalization constant $A_{\Delta_{S}}$ is defined as
 \begin{equation}
A_{\Delta_{S}} \equiv
\frac{16 \pi^{5/2}}{\left(2 \pi \right)^{2 \Delta_S}}
\frac{\Gamma \left(\Delta_{S}+1/2 \right)}
{\Gamma \left(\Delta_{S} - 1 \right)
\Gamma \left(2 \Delta_S \right)}\,.
\label{eq:PhaseSpaceVol}
 \end{equation}
 The absorptive (imaginary) part of this expression takes the form,
 \begin{equation}
\left.
\int d^4 x \; e^{ipx}
\langle 0|T\left[ \mathcal{O}_{S}^{\dagger}(x) \mathcal{O}_{S}(0)\right] |0 \rangle\right|_{\mathrm{abs.}} =
 \frac{1}{2} \frac{A_{\Delta_{S}}}
{\left(p^2\right)^{2 - \Delta_{S}} } \theta(p^2)~.
\label{georgi-norm-Im}
 \end{equation}
For $ \Delta_{S} \geq 2$, the left-hand side of Eqn.~(\ref{georgi-norm}) diverges in the ultraviolet, and we can no longer employ this normalization unless this expression is appropriately regulated. However, as shown in Appendix A, the absorptive part of Eqn.~(\ref{georgi-norm}) given in Eqn.~(\ref{georgi-norm-Im}) is ultraviolet safe and is not affected by the regulation procedure. We can therefore use Eqn.~(\ref{georgi-norm-Im}) to normalize the operator $\mathcal{O}_{S}$ over the entire range of scaling dimensions $\Delta_{S} > 1$. We shall focus on the range of scaling dimensions such that $\mathcal{O}_S$ is close to marginal, $\Delta_{S} = 4 - \epsilon$, where $\epsilon$ is small. This choice naturally allows a large hierarchy between the ultraviolet cutoff $M_{\rm UV}$ and the compositeness scale $\Lambda$.

We take the dilaton, the pseudo-Nambu-Goldstone boson associated with the spontaneous breaking of the approximate conformal symmetry, to be the lightest composite state. Then the low energy Lagrangian at the scale $\Lambda$ includes kinetic terms and mass terms for the dilaton,
 \begin{equation}
\mathcal{L}_{\rm IR} \supset
\frac{1}{2} \partial_\mu \sigma \partial^\mu \sigma - \frac{1}{2} m_{\sigma}^2 \sigma^2 \;.
 \end{equation}
Since the deformation is large at the breaking scale, $m_{\sigma}$ is parametrically of order $\Lambda$. The form of the dilaton interactions is dictated by the nonlinearly realized conformal symmetry. It is convenient to express the interactions of the dilaton in terms of the conformal compensator $\chi$, defined in terms of $\sigma$ and $f$ as 
\beq
\chi \equiv f\, e^{\sigma/f}\,.
\eeq

To realize our scenario, we require that the spectrum of light composite hidden sector states at the scale $\Lambda$ also includes three pairs of right-handed Weyl fermions $N^{\alpha}_R$ and their left-handed Dirac partners $N^{\alpha}_L$, which together form three Dirac fermions $N^\alpha=(N^{\alpha}_L, N^{\alpha}_R)$.
Here $\alpha = 1,2,3$ represents a hidden sector flavor index. The $N^\alpha$ will play the role of singlet neutrinos in the generation of neutrino masses through the inverse seesaw mechanism. We will return to the issue of flavor later in this section, but for now, we suppress all flavor indices.
In addition to kinetic terms and mass terms for these fields, the nonlinearly realized conformal symmetry gives rise to an interaction between the dilaton and the composite singlet neutrinos. The corresponding terms in the Lagrangian take the form,
 \begin{equation}
\mathcal{L}_{\rm IR} \supset
i\bar{N} \slashed{\partial} N - M_N \frac{\chi}{f} \bar N N \,.
\label{freeN}
 \end{equation}
 The mass parameter $M_N$ is of order $\Lambda$.

The CFT is assumed to couple to the SM through a neutrino portal interaction that, above the scale $\Lambda$, takes the form
 \begin{equation}
\mathcal{L}_{\rm UV} \supset -\frac{\hat{\lambda}}
{{M_{\mathrm{UV}}}^{\Delta_N - 3/2}} \bar L \widetilde H \widehat{\cal O}_{N} + {\rm h.c.}\,.
 \label{intLHO}
 \end{equation}
Here $\widehat{\cal O}_{N}$ represents a right-handed Weyl-fermionic primary operator of the CFT and $\Delta_{N}$ its scaling dimension, $L$ is the left-handed lepton doublet of the SM, and $\widetilde H \equiv i\sigma_2 H^\ast$ where $H$ is the SM Higgs doublet. The dimensionless parameter $\hat{\lambda}$ is taken to be $\mathcal{O}(1)$. 

To normalize the operator $\widehat{\cal O}_{N}$, we again employ the two-point function. Unitarity places a lower bound on the scaling dimensions of fermionic operators, $\Delta_{N} \geq 3/2$. We shall consider scaling dimensions in the range $3/2 \leq \Delta_{N} < 5/2$. For this range of scaling dimensions, we can follow the conventions of unparticle physics for fermionic operators~\cite{Cacciapaglia:2008ns},
 \begin{equation}
\int d^4 x\, e^{ipx}
\langle 0|T\left[ \widehat{\cal O}_{N}(x) \widehat{\cal O}_{N}^\dagger(0)\right] |0 \rangle =
\frac{A_{\Delta_{N} - 1/2}}{2\, i\, {\rm cos} \left( \Delta_{N} \pi \right) }
\frac{\sigma^\mu p_\mu}
{\left(-p^2 - i \epsilon\right)^{5/2 - \Delta_{N}} } \;.
 \end{equation}

 At energies of order $\Lambda$, the interaction in Eqn.~(\ref{intLHO}) gives rise to a term in the low-energy Lagrangian of the form,
 \begin{equation}
\mathcal{L}_{\rm IR} \supset 
- \lambda \left( \frac{\chi}{f} \right)^{\Delta_N - 3/2} \bar L \widetilde H N_R + {\rm h.c.}\,.
\label{intLHN}
 \end{equation}
Based on our normalization of $\widehat{\cal O}_{N}$, using the methods of ``naive dimensional analysis'' (NDA)~\cite{Weinberg:1978kz,Manohar:1983md, Georgi:1986kr} we estimate $\lambda$ to be of order
 \begin{equation}
\lambda \sim
C_\lambda\,
\hat \lambda
\left(\frac{\Lambda}{M_{\mathrm{UV}}}\right)^{\Delta_{N}-3/2}\;,
\label{lambdascaling}
 \end{equation}
where the order one multiplicative factor $C_\lambda$ is given by
 \begin{equation}
C_\lambda = \frac{(4\pi)^{3/2-\Delta_{N}}}{\Gamma(\Delta_{N}-3/2)}
\sqrt{\frac{ \pi}{ (\Delta_{N}-3/2)\cos (\Delta_{N}\pi) }} ~.
\label{lambdascaling2}
 \end{equation}
The interaction in Eqn.~(\ref{intLHN}) constitutes a coupling of the SM to the composite singlet neutrinos $N_R$ through the neutrino portal. Since $\Delta_N > 3/2$, the coupling $\lambda$ is hierarchically small for $\Lambda \ll M_{\rm UV}$.  After the Higgs acquires a vacuum expectation value (VEV), this interaction gives rise to a small mixing between the SM neutrinos and the composite singlet neutrinos $N_L$. Therefore, the light neutrinos contain a small admixture of the composite singlet states, while the heavy mass eigenstates contain a small admixture of the SM neutrino. In this way, the composite singlet neutrinos inherit a small coupling to the $W^\pm$ and $Z$ gauge bosons of the SM. 

\subsection{Neutrino Masses}

We now outline how neutrino masses can be naturally incorporated into this framework. We assume that the hidden sector possesses a global symmetry under which the operator $\mathcal{O}_N$ is charged. We can normalize the charges under the global symmetry such that $\mathcal{O}_N$, and therefore $N_R$, carries a charge of $+1$. Then, we see from Eqn.~(\ref{intLHO}) that this global symmetry can be incorporated into an overall lepton number symmetry under which both $N_L$ and $N_R$ carry charge $+1$. In order to generate Majorana neutrino masses, we require a source of lepton number violation in the theory. Accordingly, we introduce into the Lagrangian a small deformation of the CFT, denoted by $\mathcal{O}_{2N}$, which explicitly violates lepton number by two units,
 \begin{equation}
\label{Nc2deformation}
\mathcal{L}_{\rm UV} \supset -\frac{\hat{\mu}}{M_{\mathrm{UV}}^{\Delta_{2N} - 4}} \mathcal{O}_{2N} .
 \end{equation} 
 Here $\Delta_{2N}$ is the scaling dimension of the scalar operator $\mathcal{O}_{2N}$, and $\hat{\mu}$ is a dimensionless parameter. We normalize the operator $\mathcal{O}_{2N}$ based on the absorptive part of its two-point function,
 \begin{equation}
\left.
\int d^4 x\, e^{ipx}
\langle 0|T\left[ \mathcal{O}_{2N}^{\dagger}(x) \mathcal{O}_{2N}(0)\right] |0 \rangle\right|_{\mathrm{abs.}} =
 \frac{1}{2} \frac{A_{\Delta_{2N}}}
{\left(p^2\right)^{2 - \Delta_{2N}} } \theta(p^2)~.
 \end{equation}
Assuming this deformation carries a lepton number of $+2$, at scales of order $\Lambda$ it gives rise to a lepton number violating Majorana mass term for the composite singlet neutrinos,
 \begin{equation}
\mathcal{L}_{\rm IR} \supset 
-\frac{\mu}{2} \left( \frac{\chi}{f} \right)^{\Delta_{2N} - 3} \bar N^cN+\hc= -\frac{\mu}{2} \left( \frac{\chi}{f} \right)^{\Delta_{2N} - 3} \Big[\bar N^c_LN_L+\bar N^c_RN_R+\hc\Big]\,.
 \label{massN2}
 \end{equation}
 We can estimate the mass parameter $\mu$ as
 \begin{equation}
\mu\sim C_\mu \; \hat{\mu}\, \Lambda \left(\frac{\Lambda}{M_{\mathrm{UV}}}
\right)^{\Delta_{2N} - 4} ~,
\label{muscaling}
 \end{equation}
where the multiplicative factor $C_\mu$ is given by
 \bea
C_\mu &=& \frac{(4\pi)^{2-\Delta_{2N}}}{\Gamma(\Delta_{2N}-1)}
\sqrt{\frac{1}{\Delta_{2N}-1}}~.
 \eea
We see from Eqns.~(\ref{freeN}), (\ref{intLHN}) and (\ref{massN2}) that the low energy Lagrangian contains all the ingredients necessary to realize the inverse seesaw mechanism,
 \begin{equation}
 \label{L_IR}
\mathcal{L}_{\rm IR} \supset i\bar{N} {\slashed{\partial}} N- M_N\frac{\chi}{f} \bar N N- \left[
\lambda \Big(\frac{\chi}{f}\Big)^{\Delta_N-3/2} \bar L \widetilde H N_R + \frac{\mu}{2} \left(\frac{\chi}{f}\right)^{\Delta_{2N} - 3} \bar N^cN + {\rm h.c.}\right]\,.
 \end{equation}
 Integrating out $N$, we obtain a contribution to the mass of the light neutrino from the inverse seesaw mechanism,
 \begin{equation}
 	\label{m_neutrino}
{m_\nu}\mid_{{\rm inv. seesaw}} =
\left|\mu \frac{\lambda v_{\rm EW}}{M_N}\right|^2 ~,
 \end{equation}

 where $v_{\rm EW}$ represents the electroweak vacuum expectation value (VEV) of the SM Higgs, $v_{\rm EW} \approx 174$ GeV.  We expect comparable but somewhat smaller contributions to the neutrino mass from integrating out the higher mass singlet fermion resonances, and so this expression is only an estimate.

\subsection{The Neutrino Portal}

 After electroweak symmetry breaking, the neutrino portal interaction in Eqn.~(\ref{intLHN}) and the mass term in Eqn.~(\ref{freeN}) give rise to mixing between the SM neutrinos and the singlet neutrinos. We define 
 \beq
 \label{U_NL}
 U_{N\ell} \equiv \lambda \frac{v_{\rm EW}}{M_N} \;.
 \eeq 
 This allows the expression for the neutrino mass to be written as $m_\nu = |\mu U_{N\ell}|^2$. Since $|\mu| \lesssim \Lambda$, the mixing angle $|U_{N\ell}|$ is constrained to lie in the range
  \beq
  \sqrt{\frac{m_\nu}{M_N}} \lesssim |U_{N\ell}| \lesssim 1\,.
  \label{U_Nl_range}
  \eeq
 It follows from this that the heavier mass eigenstate, which we continue to denote as $N$, inherits couplings to the $W$ and $Z$ gauge bosons from the SM neutrino it mixes with, suppressed by a factor of the mixing parameter $U_{N\ell}$,
 \beq
\mathcal{L}_{\rm IR} \supset g_Z U_{N\ell} [\overline{\nu_\ell} Z_\mu \gamma^\mu P_L N] + g_W U_{N\ell} [\overline{\ell} W^{-}_\mu \gamma^\mu P_L N] + \hc \,.
 \label{eq:effNcoupling}
 \eeq
The mixing between the SM neutrinos and the singlet neutrinos also gives rise to couplings between the light neutrino mass eigenstate, which we continue to denote as $\nu$, and the composite states $N$ and $\sigma$. Expanding out the dilaton interactions in Eqns.~(\ref{freeN}) and (\ref{intLHN}) to leading order in $1/f$, we have
\beq
\mathcal{L}_{\rm IR} \supset -\lambda (\Delta_N - 3/2) \bar L\widetilde H N_R \frac{\sigma}{f} - M_N \frac{\sigma}{f} \bar N_L N_R +\hc\,.
\eeq
Going over to the mass eigenstate basis, after electroweak symmetry breaking, we have
\begin{align}
\mathcal{L}_{\rm IR} &\supset -(\Delta_N - 3/2)  \frac{M_N}{f} U_{N\ell}\, \bar \nu_L N_R \sigma + \frac{M_N}{f} U_{N\ell}\, \bar \nu_L N_R \sigma + \hc	\notag\\
 &=-(\Delta_N - 5/2)  \frac{M_N}{f} U_{N\ell}\, \bar \nu_L N_R \sigma + \hc \,.
\label{eq:Nnusigma}
\end{align}
In the first line above, we have employed Eqn.~(\ref{U_NL}). 
This interaction allows an $N$ to decay into a $\sigma$ and a light neutrino. This is the dominant decay mode of the singlet neutrinos in a large part of the parameter space relevant for current and near-future experiments.

After mixing, the lepton number violating interaction in \eqref{massN2} also gives rise to a coupling between the dilaton $\sigma$ and a light neutrino and antineutrino. To leading order in $1/f$, we have
\begin{align}
\mathcal{L}_{\rm IR} &\supset -(2\Delta_N + \Delta_{2N} - 8)  \frac{\mu}{2f} U^2_{N\ell}\, \bar \nu_L \nu^c_L \sigma + \hc \notag\\
&=-(2\Delta_N + \Delta_{2N} - 8) \frac{m_\nu}{2f}\,  \bar \nu_L \nu^c_L \sigma + \hc,
\label{eq:nunusigma}
\end{align}

where we have used Eqn.~(\ref{m_neutrino}). This interaction allows the dilaton to decay into a neutrino and antineutrino. This is the primary decay mode of the dilaton in some part of the parameter space relevant for current and near-future experiments. 

We now return to the realistic case of three generations of SM fermions and singlet neutrinos. Then, the Dirac mass parameter $m_N$ for the singlet neutrinos in Eqn.~(\ref{freeN}), the couplings $\hat{\lambda}$ and $\lambda$ in Eqns.~(\ref{intLHO}) and (\ref{intLHN}), and the lepton number violating parameters $\hat{\mu}$ and $\mu$ in Eqns.~(\ref{Nc2deformation}) and (\ref{massN2}), all generalize to $(3 \times 3)$ matrices in flavor space, $m_{N, \alpha \beta}$, $\hat{\lambda}_{i \alpha}$, $\lambda_{i \alpha}$, $\hat{\mu}_{\alpha \beta}$ and $\mu_{\alpha \beta}$ respectively. Here $i=1,2,3$ represents a flavor index for the SM leptons $L^i$ and $\alpha =1,2,3$ a flavor index for the singlet neutrinos $N^\alpha$. For simplicity, we will assume that the singlet neutrinos $N^\alpha$ are approximately degenerate in mass and that the couplings $\hat{\lambda}_{i \alpha}$ and $\lambda_{i \alpha}$ are flavor diagonal and universal, so that $m_{N, \alpha \beta} \; \propto \; \delta_{\alpha \beta}$,
 $\hat{\lambda}_{i \alpha} \; \propto \; \delta_{i \alpha}$, $\lambda_{i \alpha} \; \propto \; \delta_{i \alpha}$.{\footnote{Deviations from the flavor-diagonal form of these couplings will, in general, lead to lepton flavor-violating processes. The rates for such processes, and the resulting constraints on off-diagonal couplings in flavor space were discussed in Ref.~\cite{Chacko:2020zze}.}} Then, in this framework. the nontrivial flavor structure of the mass matrix for the light neutrinos arises from the lepton number violating  parameters, $\hat{\mu}_{\alpha \beta}$ and $\mu_{\alpha \beta}$, which are not taken to be flavor universal.

 Under this assumption that couplings of the SM to the hidden sector are flavor universal, any lepton flavor violating process requires at least one insertion of $\mu$, and is therefore suppressed by the light neutrino masses. This construction is therefore naturally consistent with the existing constraints on lepton flavor violation. Furthermore, tests of lepton universality do not constrain the parameter space. Specifically, ratios of the decay widths  of mesons into different flavors of charged leptons, such as those shown in Table~1 of \cite{deGouvea:2015euy}, are not sensitive to a hidden sector that couples flavor universally to the SM.

\subsection{The Higgs Portal}

At loop order, Eqns.~(\ref{freeN}) and (\ref{intLHN}) also generate a coupling of the Higgs field to the dilaton,
\begin{align}
\label{HPortal2}
 \mathcal{L}_{\rm IR} &\supset -g_N \Lambda^2 H^{\dagger} H \left(\frac{\chi}{f}\right)^{2 \Delta_N - 1} \;.
\end{align}
 Here $g_N$ is of order $3\lambda^2/(16 \pi^2)$, where the factor of 3 arises from the three flavors of leptons.
In general, we also expect a direct Higgs portal coupling of the hidden sector to the SM in the ultraviolet of the form,
\begin{align}
\mathcal{L}_{\rm UV} &\supset  -\frac{\hat{g}}{M_{\mathrm{UV}}^{\Delta_S - 2}} H^{\dagger} H \mathcal{O}_S \;. \label{HPortal1}
\end{align}
 In this expression, the mass scale $M_{\mathrm{UV}}$ represents the 
ultraviolet cutoff of the theory, and $\hat{g}$ is a dimensionless 
parameter taken to be $\mathcal{O}(1)$. 
If the three-point function $\langle \mathcal{O}_N(x) 
\mathcal{O}_N^{\dagger}(y) \mathcal{O}_S(z)\rangle$ is non-vanishing, quantum effects will give rise to a contribution to the Higgs portal coupling in Eqn.~(\ref{HPortal1})  from the neutrino portal interaction in Eqn.~(\ref{intLHO}). For the range of scaling dimensions such that $\Delta_S>2\Delta_N-1$, this effect is infrared-dominated and constitutes the leading contribution to $\hat{g}$ at low energies. Since we are focusing on the range of scaling dimensions for which $3/2 < \Delta_N < 5/2$ and $\Delta_S = 4 -\epsilon$, the condition $\Delta_S>2\Delta_N-1$ is satisfied in most of the parameter space.
The size of this contribution to $\hat{g}$ at an arbitrary scale $\Lambda_{\rm UV} < M_{\mathrm{UV}}$ can be obtained by integrating out modes with momenta $p$ in the range $M_{\rm UV} > p > \Lambda_{\rm UV}$. By matching to the Higgs portal interaction in Eqn.~(\ref{HPortal2}) at the scale $\Lambda_{\rm UV} = \Lambda$, we can estimate this contribution to $\hat{g}$ for arbitrary $\Lambda_{\rm UV}$ as, 
 \begin{equation}\label{matching_a}
\hat{g} \sim \frac{3\hat{\lambda}^2}{16\pi^2} C_{\lambda}^2 \sqrt{\frac{16\pi^3}{A_{\Delta_S}}} \left(\frac{M_{\rm UV}}{\Lambda_{\rm UV}}\right)^{(\Delta_S - 2 \Delta_N + 1)} \;.
 \end{equation}
 For the range of scaling dimensions $\Delta_S < 2 \Delta_N -1 $,  the radiative contribution of the neutrino portal interaction in Eqn.~(\ref{intLHO}) to $\hat{g}$ is subdominant and can be neglected. For this range of scaling dimensions, at low energies, the interaction in Eqn.~(\ref{HPortal1}) gives rise to a coupling of the Higgs to the dilaton of the form,
\begin{align}
\mathcal{L}_{\rm IR} &\supset - g_S \Lambda^2 H^{\dagger} H \left(\frac{\chi}{f}\right)^{\Delta_S} \;.
\label{HPortal3}
\end{align}
 Based on our normalization of $\mathcal{O}_S$, we can estimate the value 
of $g_S$,
 \begin{align}
 g_S \sim \hat{g}\sqrt{\frac{A_{\Delta_S}}{16\pi^3}}
\left(\frac{\Lambda}{M_{\mathrm{UV}}}\right)^{\Delta_S - 2}. 
 \end{align}

The Higgs can directly decay into the hidden sector through the Higgs portal interaction in Eqn.~(\ref{HPortal1}). For a general Higgs portal coupling $\hat{g}$, the inclusive decay 
width is given by
 \begin{equation}\label{higgs_decay_inclusive}
\Gamma_{h\rightarrow \mathcal{O}_S} = A_{\Delta_S}\frac{2\hat{g}^2v_{\rm EW}^2}{m_h}\left(\frac{m_h^2}{M_{\rm UV}^2}\right)^{\Delta_S-2}\,.
 \end{equation}
 For the range of scaling dimensions $\Delta_S>2\Delta_N-1 $, the appropriate value of $\hat{g}$ is to be obtained from Eqn.~(\ref{matching_a}) after setting $\Lambda_{\rm UV}=m_h$. For $\Delta_S < 2\Delta_N-1$, the appropriate choice of $\hat{g}$ is the value at the ultraviolet scale $M_{\rm UV}$. In both cases, the suppression of the rate by a sizable power of $m_h^2/M_{\rm UV}^2$ means that the effect of the Higgs portal coupling in Eqn.~(\ref{HPortal1}) on Higgs decays is small and can be neglected. 
 
After expanding out $\chi$ in terms of $\sigma$, the operators Eqns.~(\ref{HPortal2}) and (\ref{HPortal3}) lead to mixing between the dilaton and the Higgs boson,
\beq
\label{HPortal4}
{\cal L}\supset - g\Lambda^2 \frac{v}{f}\, \sigma h \;, 
\eeq 
where, for convenience, we have defined $g \equiv g_N(2\Delta_N - 1)$ for $\Delta_S>2\Delta_N-1$  and $ g \equiv g_S (\Delta_S - 2)$ for $\Delta_S<2\Delta_N-1 $. Taking $m_\sigma^2 \ll m_h^2$, we can diagonalize the mass matrix and find the mass eigenstates. The corresponding mixing angle is given by

\beq
\sin \theta \equiv \left|\frac{g \Lambda^2 v_{\rm EW}}{f m_h^2}\right|.
\label{sintheta}
\eeq
For most of the parameter space, $\Delta_S>2\Delta_N-1 $, and so we have $g \approx g_N (2\Delta_N - 1)$. Then $\sin \theta$ can be estimated as
\beq
\sin \theta \sim \left|\frac{g_N \Lambda^2 v_{\rm EW}}{f m^2_h}\right| \sim |U_{N\ell}|^2\left(\frac{m_N}{v_{\rm EW}}\right)\left(\frac{m_N}{m_h}\right)^2 \;.
\label{sintheta2}
\eeq

Correspondingly, $\sigma$ inherits couplings to SM states through the Higgs, suppressed by a factor of $\sin \theta$. However, as we show in the next section, this mixing has no significant effect on the production and decay of $\sigma$ particles in the parameter space of interest. Therefore, it will play no role in the phenomenology of this class of models.

\section{Signals at Colliders and Beam Dumps}

In this section, we consider the signals of this class of models at colliders and beam dumps. We determine the current constraints on the parameter space of these theories and explore the reach of future searches. We choose the parameters $m_\sigma/m_N = 0.6$, $\Delta_{2N} = 17/4$ and $\Delta_N = 7/4$ as benchmark values for this discussion. For comparison, we will later consider the values $\Delta_N = 1.9$ and $\Delta_N = 2.1$ as well. We will focus on compositeness scales $\Lambda \gtrsim$ 100 MeV, for which the cosmological constraints will be seen to be naturally satisfied. Taking $m_N \sim \Lambda \sim 4\pi f$, the $\sigma$-$\nu$-$N$ coupling shown in Eqn.~(\ref{eq:Nnusigma}) can be written as,
 \beq
|g_{\sigma \nu N}| \approx 2\pi \left(2\Delta_N - 5\right) U_{N\ell}.
 \eeq
Similarly, the $\sigma$-$\nu$-$\nu^c$ coupling shown in Eqn.~(\ref{eq:nunusigma}) can be expressed as,
\beq
|g_{\sigma \nu \nu}| \approx 2\pi \left(2\Delta_N + \Delta_{2N} - 8\right) \frac{m_\nu}{m_N}.
\eeq
Although these expressions for the couplings are only approximate, we shall treat them as exact for the purposes of our analysis. This does not qualitatively affect our results.

\bigskip

\subsection{Production and Decay of Hidden Sector States}
\label{subsec: prod_and_decay}


The hidden sector couples to the SM through the neutrino portal interaction shown in Eqn.~(\ref{intLHO}). At colliders and beam dumps, this interaction allows hidden sector particles to be produced through the decays of an on- or off-shell $W$, $Z$, or Higgs boson. Since the hidden sector is strongly coupled, the operator ${\mathcal O}_{N}$ can give rise to multiparticle states. However Lorentz invariance and angular momentum conservation place restrictions on the possible states that can be produced in this way. In particular, since $\mathcal{O}_N$ is a fermion, the final state particles it generates must always contain at least one $N$. More generally, states of the form $n\, N + (n-1)\,\overline{N}+p\,\sigma$ are allowed for $n\ge 1$ and $p\ge 0$. Additional $\sigma$ particles can be produced from the subsequent decays of $N$ and $\overline{N}$, as discussed below. While the possibility of such multiparticle states is interesting in its own right and can give rise to exotic signatures, for the parameter region of interest, the majority of events involve the production of just a single $N$, sometimes accompanied by one or two additional dilatons $\sigma$. These hidden sector states are produced in association with a charged lepton, if produced through a $W$ boson, or with a neutrino, if produced through a $Z$ or Higgs boson. The dominant production channel depends on the compositeness scale, with two distinct possibilities:
\begin{enumerate}
    \item [(i)] For values of $m_N$ below a few GeV, production is dominated by leptonic and semileptonic decays of mesons, as these are produced in abundance through the strong interactions. In the mass range $m_N \lesssim 350~\text{MeV}$, there are severe constraints on this scenario from pion and kaon decays. For $m_N > 350~\text{MeV}$, the decays of charm and bottom mesons such as $D^\pm, D^0, D^\pm_S, B^0$, and $B^\pm$ play the dominant role. 
    
    
    The branching fractions of all the relevant channels (for the case of an HNL that couples to electrons) were calculated in Ref.~\cite{Bondarenko:2018ptm}, and are shown in Fig.~\ref{fig: meson_br}. In our case, due to the unparticle phase space in the final state, the decay widths are modified as compared to the case of a ``conventional" (weakly coupled, narrow) HNL. For decays of the form $X \rightarrow \mathcal{U}_N Y$ ($X$ here represents a SM particle, $Y$ a system of up to two SM particles, and $\mathcal{U}_N$ an arbitrary hidden sector state), the decay widths can be calculated by appropriately modifying the known decay widths in the case of conventional HNLs. Denoting the decay width of $X \rightarrow \widetilde{N} Y$ by $\widetilde{\Gamma}(m_\mathcal{U})$, where $\widetilde{N}$ is a conventional HNL of mass $m_\mathcal{U}$, the differential decay width with respect to the continuous mass of $\mathcal{U}_N$ is given by,
    \begin{equation}
        \frac{d\Gamma (X \rightarrow \mathcal{U}_N Y)}{dm^2_\mathcal{U}} = \frac{\mathcal{A}_{\Delta_N - \frac{1}{2}}}{2 \pi C^2_\lambda m^2_\mathcal{U}} \left(\frac{m^2_\mathcal{U}}{\Lambda^2}-1\right)^{\Delta_N - \frac{5}{2}} \widetilde{\Gamma} \left(m_\mathcal{U}\right) \;,
    \label{eq: meson_two_body_width}
    \end{equation}
    where $m^2_\mathcal{U} \in \left[\Lambda^2, (m_X - m_Y)^2\right]$. A derivation of Eqn.~(\ref{eq: meson_two_body_width}) may be found in Appendix~\ref{app: uN_prod}. The matrix elements for leptonic and semi-leptonic decay modes of pseudoscalar mesons that contain conventional HNLs in the final state may be found in Appendix~\ref{app: matel}. Using the differential decay width shown in Eqn.~(\ref{eq: meson_two_body_width}), we perform a Monte Carlo analysis to sample events and to determine the partial decay widths of each channel. Details of our Monte Carlo analysis may be found in Appendix~\ref{app: MC}. 
    
    \begin{figure}[h!]
    \centering
    \begin{subfigure}{0.41\textwidth}
        \includegraphics[width=\linewidth]{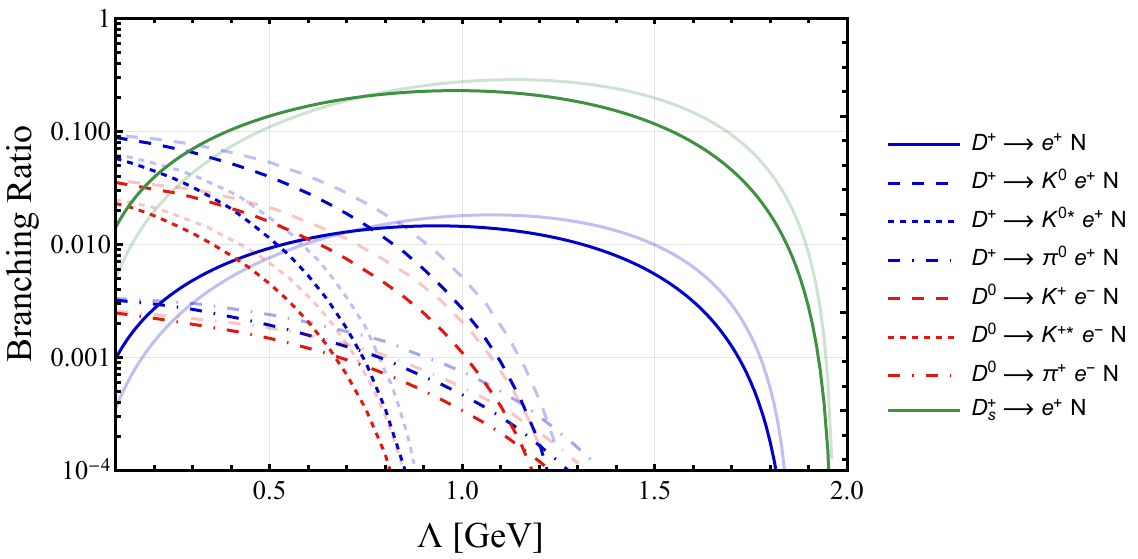}
    \end{subfigure}
    \hfill
    \begin{subfigure}{0.49\textwidth}
        \includegraphics[width=\linewidth]{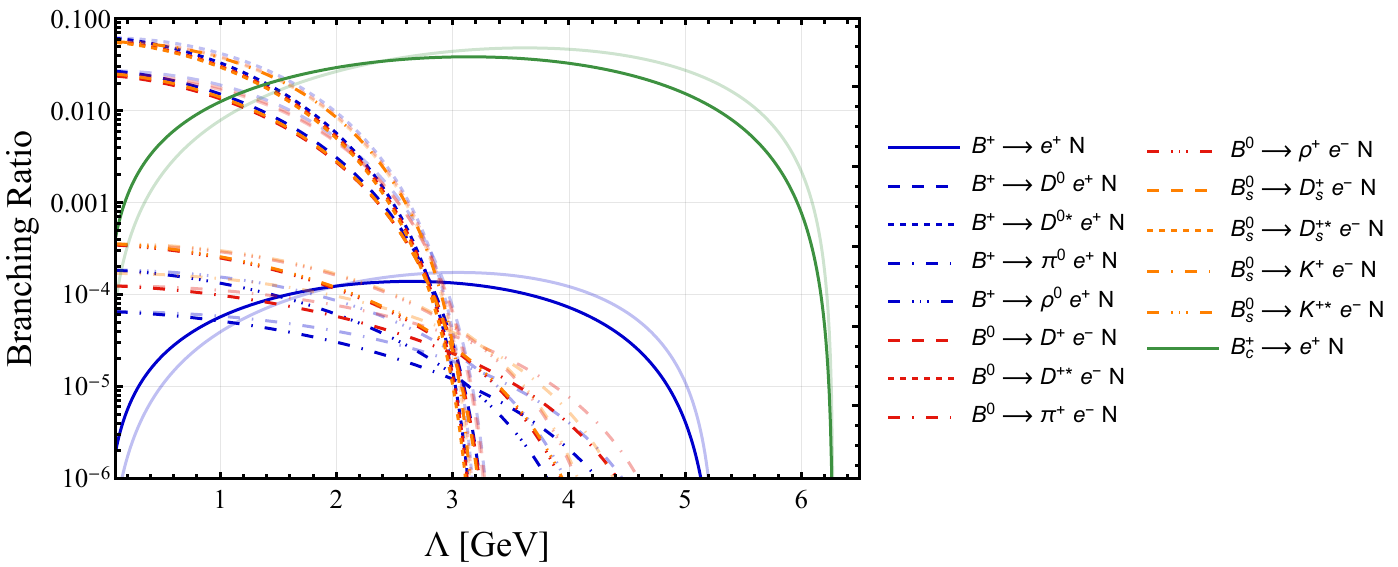}
    \end{subfigure}
    \bigskip
    \caption{Branching fractions for heavy meson decays $D \rightarrow \mathcal{U}_N X$ (left) and $B \rightarrow \mathcal{U}_N X$ (right) as a function of $\Lambda$. These branching ratios are calculated using Eqn.~(\ref{eq: meson_two_body_width}) and correspond to the case where $\Delta_N = 7/4$ and $|U_{N\ell}|^2 = 1$. The corresponding branching ratios for the case of a conventional HNL, taken from~\cite{Bondarenko:2018ptm}, are shown in lighter shades.}
    \label{fig: meson_br}
    \end{figure} 

    \item [(ii)] For heavier masses, $m_N \gtrsim m_{B_C} \sim 6~\text{GeV}$, the dominant production mechanism for hidden sector states is through the decays of $W^\pm$ and $Z$ bosons at high energy colliders.\footnote{For large values of $\Delta_N$, production through decays of the Higgs boson may also play a significant role~\cite{Borrello:2025hal}. However, this channel is less important for the range of $\Delta_N$ that we are considering.} It was shown in Ref.~\cite{Chacko:2020zze} that in the range of $\Delta_N$ that we are considering, the decays of $W$ to a single $N$ are preferred over multiple $N$. When produced from a $W^\pm$, the $N$ is accompanied by a charged lepton. We consider this production channel in Section~3.5.4 in the context of displaced vertex searches for HNLs at the LHC. The decay width for $W^\pm \rightarrow \ell \widetilde{N}$, where $\widetilde{N}$ is a conventional HNL of mass $m_\mathcal{U}$, is straightforward to calculate as a function of the mixing angle. The result is given (in the $m_{\ell}=0$ limit) by
    \begin{equation}
        \widetilde{\Gamma}(m_\mathcal{U}) = \frac{g^2 m_W}{48 \pi}|U_{N\ell}|^2 \left(1 - \frac{m^2_\mathcal{U}}{m^2_W}\right)^2 \left(1 + \frac{m^2_\mathcal{U}}{2m^2_W}\right).
    \end{equation}
     We can then use Eqn.~(\ref{eq: meson_two_body_width}) to calculate the corresponding decay widths in our scenario.
    
\end{enumerate}

\begin{figure}[h!]
    \centering
    \includegraphics[width=0.7\linewidth]{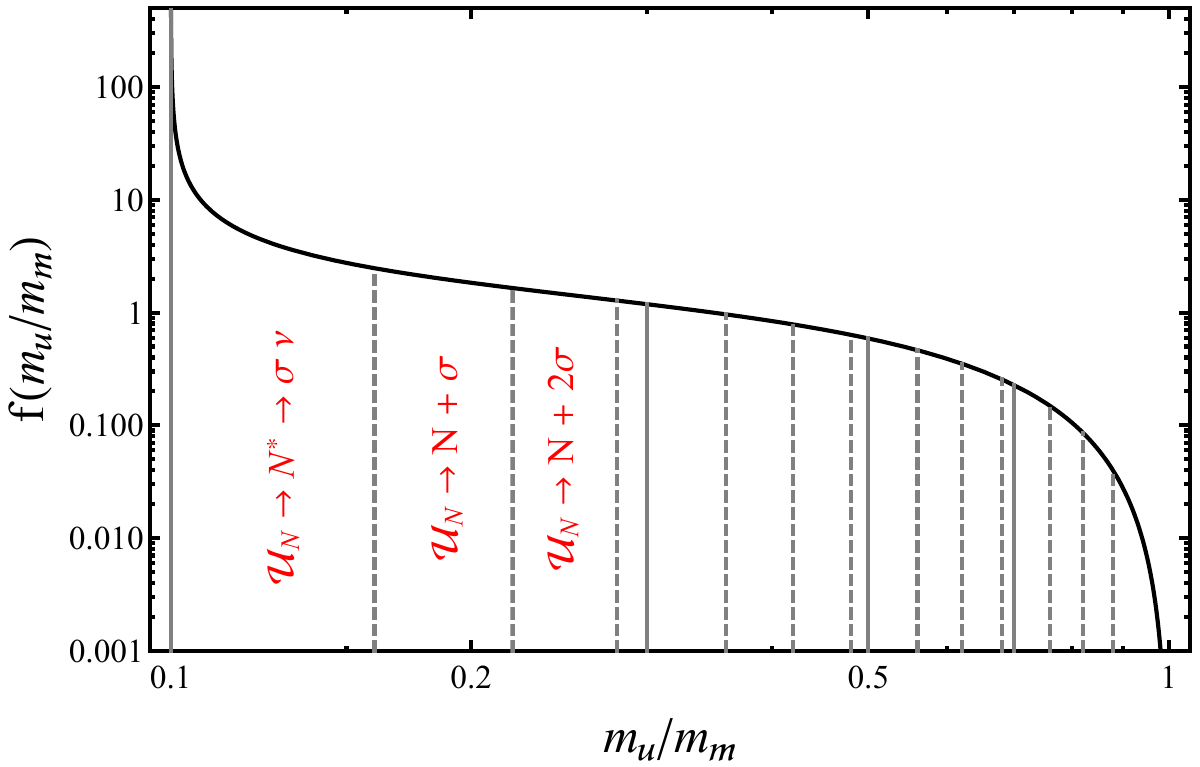}
    \caption{Probability density function of $m_\mathcal{U}$ in the case of leptonic decays of pseudoscalar mesons, where we set $m_\ell=0, ~\Delta_N = 7/4, ~m_\sigma/m_N = 0.6$ and $\Lambda/m_\mathfrak{m} = 0.1$. As shown here, we divide the kinematically allowed range in $m_\mathcal{U}$ into intervals corresponding to final states of the form $n\, N + (n-1)\,\overline{N}+p\,\sigma$. Solid lines correspond to $n=1, 2, 3, \ldots$ and the dashed lines correspond to $p=1, 2, 3$.}
    \label{fig: unparticle_to_composites}
\end{figure}

As discussed earlier, for a given production channel $X \rightarrow \mathcal{U}_N Y$, we are ultimately interested in final states containing the composite states of the hidden sector, which are of the form $n\, N + (n-1)\,\overline{N}+p\,\sigma$. Using the distribution in $m_\mathcal{U}$ obtained from Eqn.~(\ref{eq: meson_two_body_width}), we bin the kinematically allowed range of $m_\mathcal{U}$ using the edges $(m_N, m_N + m_\sigma, \cdots, m_N + km_\sigma, 3m_N, \cdots ,m_\mathrm{max})$, where $k = \lfloor 2m_N/m_\sigma \rfloor$. In each interval, we assume that the final state with the largest sum of masses of its constituent composite states that is kinematically allowed is the dominant one. For example, in the interval $m_\mathcal{U} \in [m_N + m_\sigma, m_N + 2m_\sigma]$, this would be $N + \sigma$. Moreover, as we show below, the composite state $N$ promptly decays into $\sigma + \nu$. As a result, in the interval $m_\mathcal{U} \in [m_N, m_N + m_\sigma]$, we take $m_\mathcal{U}$ to be the invariant mass of the $\sigma + \nu$ system. Our treatment of the final states containing composite particles of the hidden sector is illustrated for the case of leptonic decays of pseudoscalar mesons in Fig.~\ref{fig: unparticle_to_composites}.

We now turn our attention to how the hidden sector particles decay. The composite singlet neutrinos $N$ can decay into a dilaton and a neutrino, $N \rightarrow \sigma \nu$, through the interactions in Eqn.~(\ref{eq:Nnusigma}), or directly into SM states through their couplings to the $W$ and $Z$ gauge bosons in Eqn.~(\ref{eq:effNcoupling}). Although both these interactions are suppressed by a single factor of the mixing angle $U_{N\ell}$, the latter is further suppressed by a factor $ \sim (m_N/v_{\rm EW})^4$ as well as by the three-body phase space factor. Therefore, the $N \rightarrow \sigma \nu$ mode dominates in our mass range of interest ($m_{N}\ll v_{\rm EW}$). For the $N \rightarrow \sigma \nu$ channel, the decay width is given by 
 \begin{equation}
    \Gamma_N = \frac{\pi m_N}{8} \left(2\Delta_N - 5\right)^2|U_{N\ell}|^2 \left(1 - \frac{m^2_\sigma}{m^2_N}\right)^2.
 \end{equation}
It follows from this that $N$ decays promptly in the entire parameter range of interest. For reference, for $m_N = 1~\text{GeV}$ and $|U_{Nl}|^2 = 10^{-10}$, $c\tau_N$ is of the order $10^{-5}~$m. 

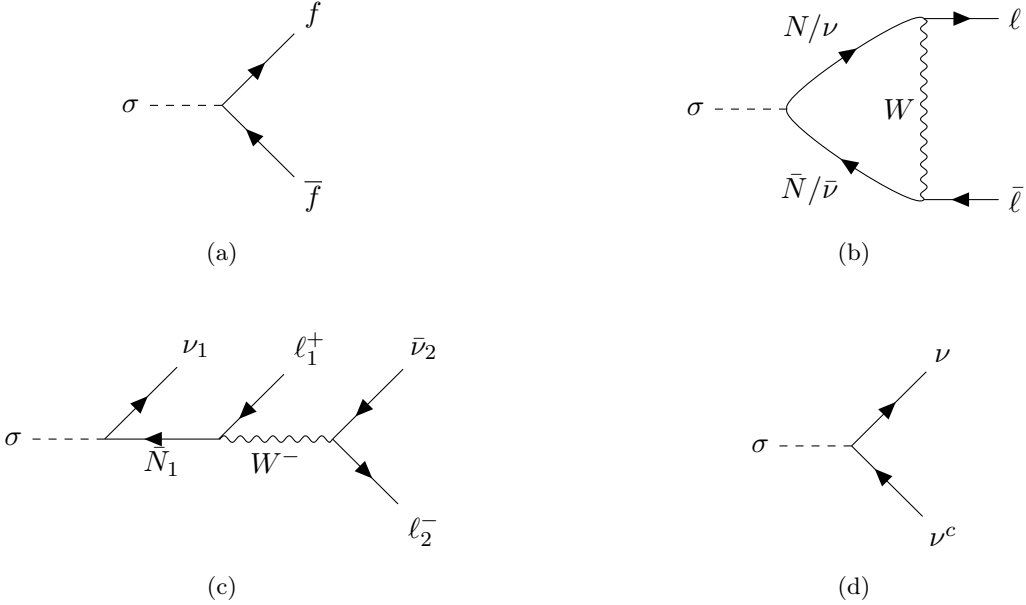
\begin{figure}[h!]
  \centering
  \begin{subfigure}{0.45\textwidth}
    \centering
    \begin{tikzpicture}[scale=0.6]
        \begin{feynman}
            \vertex (i) at (-3, 0) {\(\sigma\)};
            \vertex (a) at (-1, 0);
            \vertex (f1) at (1, 2) {\(f\)};
            \vertex (f2) at (1, -2) {\(\overline{f}\)};
            \diagram*{
            (i) -- [scalar] (a),
            (a) -- [fermion] (f1),
            (a) -- [anti fermion] (f2),
            };
        \end{feynman}
    \end{tikzpicture}
    \caption{}
    \label{fig: sigma_2_1}
  \end{subfigure}
  \hfill
  \begin{subfigure}{0.45\textwidth}
    \centering
        \begin{tikzpicture}[scale= 0.6]
        \begin{feynman}
        \vertex (a) at (-4,0){$\sigma$};
        \vertex (i1) at (-2, 0);
        \vertex (i4) at (1,-2);
        \vertex(i5) at (1, 2);
        \vertex (b) at (3, -2){$\bar{\ell}$};
        \vertex (c) at (3, 2){${\ell}$};
        \diagram*{
        (a)--[scalar](i1),
        (i4)--[fermion,edge label = $\bar{N}/\bar{\nu}$, half left, looseness=0.25](i1),
        (i1)--[fermion,edge label = ${N/\nu}$, half left, looseness=0.25](i5),
        (b)--[fermion](i4),
        (i5)--[fermion](c),
        (i4)--[boson,edge label = $W$, inner sep = 3pt](i5),
        };
        \end{feynman};
        \end{tikzpicture}
    \caption{}
    \label{fig: sigma_2_2}
    \end{subfigure}
  \par\vspace{2pc}
  \begin{subfigure}{0.45\textwidth}
    \centering
    \begin{tikzpicture}[scale=0.6]
    \begin{feynman}
      \vertex (i1) at (-3,  0) {\(\sigma\)};
      \vertex (a)  at (-1, 0);
      \vertex (f1) at ( 1,  2) {\(\nu_{1}\)};
      \vertex (b)  at ( 1.5,  0);
      \vertex (f2) at ( 3.5,  2) {\(\ell^{+}_{1}\)};
      \vertex (c)  at ( 4, 0);
      \vertex (f3) at ( 6, 2) {\(\bar{\nu}_{2}\)};
      \vertex (f4) at ( 6, -2) {\(\ell^{-}_{2}\)};
      \diagram*{
        (i1) -- [scalar] (a),
        (a)  -- [fermion] (f1),
        (a)  -- [anti fermion, edge label'={\(\bar{N}_{1}\)}, inner sep = 3pt]  (b)
              -- [anti fermion]  (f2),
        (b)  -- [boson, edge label'={\(W^{-}\)}] (c),
        (c)  -- [anti fermion] (f3),
        (c)  -- [fermion] (f4),
      };
    \end{feynman}
    \end{tikzpicture}
    \caption{}
    \label{fig: sigma_4}
  \end{subfigure}
  \hfill
  \begin{subfigure}{0.45\textwidth}
    \centering
    \begin{tikzpicture}[scale=0.6]
    \begin{feynman}
      \vertex (i1) at (-3, 0) {\(\sigma\)};
      \vertex (a)  at (-1, 0);
      \vertex (f1) at ( 1, 2) {\(\nu\)};
      \vertex (f2) at ( 1, -2) {\(\nu^c\)};
      \diagram*{
        (i1) -- [scalar] (a),
        (a)  -- [fermion] (f1),
        (a)  -- [anti fermion] (f2),
      };
    \end{feynman}
    \end{tikzpicture}
    \caption{}
    \label{fig: sigma_2_3}
  \end{subfigure}
  \par\vspace{2pc}
  \caption{Diagrams contributing to decays of $\sigma$ at order $U^2_{N\ell}$ in the amplitude. As described in the text, decay modes corresponding to~\ref{fig: sigma_2_1} and~\ref{fig: sigma_2_2} are suppressed by factors of $m^2_f$ and $m^2_\ell$ respectively. Consequently, $\sigma$ primarily decays through one of the decay modes shown in~\ref{fig: sigma_4} and~\ref{fig: sigma_2_3}.}
  \label{fig: sigma_decays_1}
\end{figure}

Since $\sigma$ is the lightest state in the hidden sector, it must decay back to the SM. The decay modes of $\sigma$ all involve at least two factors of the mixing angle $U_{N\ell}$ at the amplitude level, which means that it is long-lived. Let us survey the most likely $\sigma$ decay modes (those with only two factors of $U_{N\ell}$) in order to determine the dominant channels.
\begin{enumerate}
    \item [(i)] Through its mixing with the Higgs, $\sigma$ has two body decay modes into pairs of SM fermions, as shown in Fig.~\ref{fig: sigma_2_1}. Since $\sin \theta$ is proportional to $|U_{N\ell}|^2$, this carries two factors of $U_{N\ell}$ at the amplitude level. The partial decay widths  of these modes are given by
    \beq
    \Gamma^{(2, h)} = \frac{m_\sigma}{8\pi}  \sin^2 \theta \left(\frac{m_f}{v_{\rm EW}}\right)^2 \left(1 - \frac{4m^2_f}{m^2_\sigma}\right)^{3/2} \;.
    \eeq
    Since the Higgs is involved, this mode is suppressed by a Yukawa coupling at the amplitude level, which explains the factor of $m^2_f$.

    \item [(ii)] The dilaton $\sigma$ can also decay into a pair of leptons through a one-loop process, as shown in Fig.~\ref{fig: sigma_2_2}. The partial decay widths  of these channels are given by~\cite{Airen:2025uhy},
    \begin{equation}
        \Gamma^{(2, W)} = \frac{m_\sigma}{256\pi^3}  \left(2\Delta_N - 5\right)^2|U_{Nl}|^4 \left(\frac{m_N m_{\ell}}{v^2_{\rm EW}}\right)^2 \left(1 - \frac{4 m^2_\ell}{m^2_\sigma}\right)^{1/2} \;.
        \label{eq:sigmawidth} 
    \end{equation}
    Note that this mode is also suppressed by $m_{\ell}^2$. This is a consequence of angular momentum conservation and the chiral coupling of the $W$-bosons.

    \item [(iii)] The four-body decays shown in Fig.~\ref{fig: sigma_4} are not suppressed by angular momentum considerations or small Yukawa couplings and represent the dominant decay mode for most of the parameter space of interest, in spite of the four-body phase space suppression. The tree-level decay width can be estimated as
    \begin{equation}
        \Gamma^{(4)} \sim \frac{m_\sigma}{512 \pi^3} \left(2\Delta_N - 5\right)^2 |U_{N\ell}|^4 \left(\frac{m_\sigma}{v_{\rm EW}}\right)^4 \left(\frac{m_\sigma}{m_N}\right)^4 \;.
    \end{equation}
    A precise calculation of the width requires the evaluation of a complicated phase space integral. For values of $m_\sigma \gtrsim 1~\mathrm{GeV}$, we compute the decay widths  numerically using \texttt{MadGraph} by considering all the parton-level final states such as $\ell\ell\nu\nu, \ell\nu q q, \nu\nu q q, \nu\nu\nu\nu$, etc. For $m_\sigma \lesssim 1~\mathrm{GeV}$, we consider final states containing up to four particles, of which no more than two are pseudo-scalar mesons $\mathfrak{m}$. This includes modes such as $\ell\nu\mathfrak{m}$ and $\ell\nu\mathfrak{m}\mathfrak{m}$. 

    \item [(iv)] From the $\sigma-\nu-\nu^c$ coupling in Eqn.~(\ref{eq:nunusigma}), $\sigma$ can also decay directly to two light neutrinos. This is shown in Fig.~\ref{fig: sigma_2_3}. Since this process violates lepton number and is also forbidden by angular momentum considerations in the $m_{\nu}=0$ limit, the rate is suppressed by $m_\nu/m_N$ at the amplitude level, leading to
    \begin{equation}
    \begin{split}
        \Gamma^{(2, \nu)} = \frac{\pi m_\sigma }{2} \left(2\Delta_N - 5\right)^2\left(\frac{m_\nu}{m_N}\right)^2 \;.
    \end{split}
    \end{equation}
    Despite this large suppression factor, we find that this two-body decay mode can dominate in the part of the parameter space where $m_N |U_{N\ell}|^2 < m_\nu$ (we take $m_{\nu} \sim 0.01$ eV). 
    
\end{enumerate}

\begin{figure}[h!]
    \centering
    \begin{subfigure}{0.45\textwidth}
        \includegraphics[width=\linewidth]{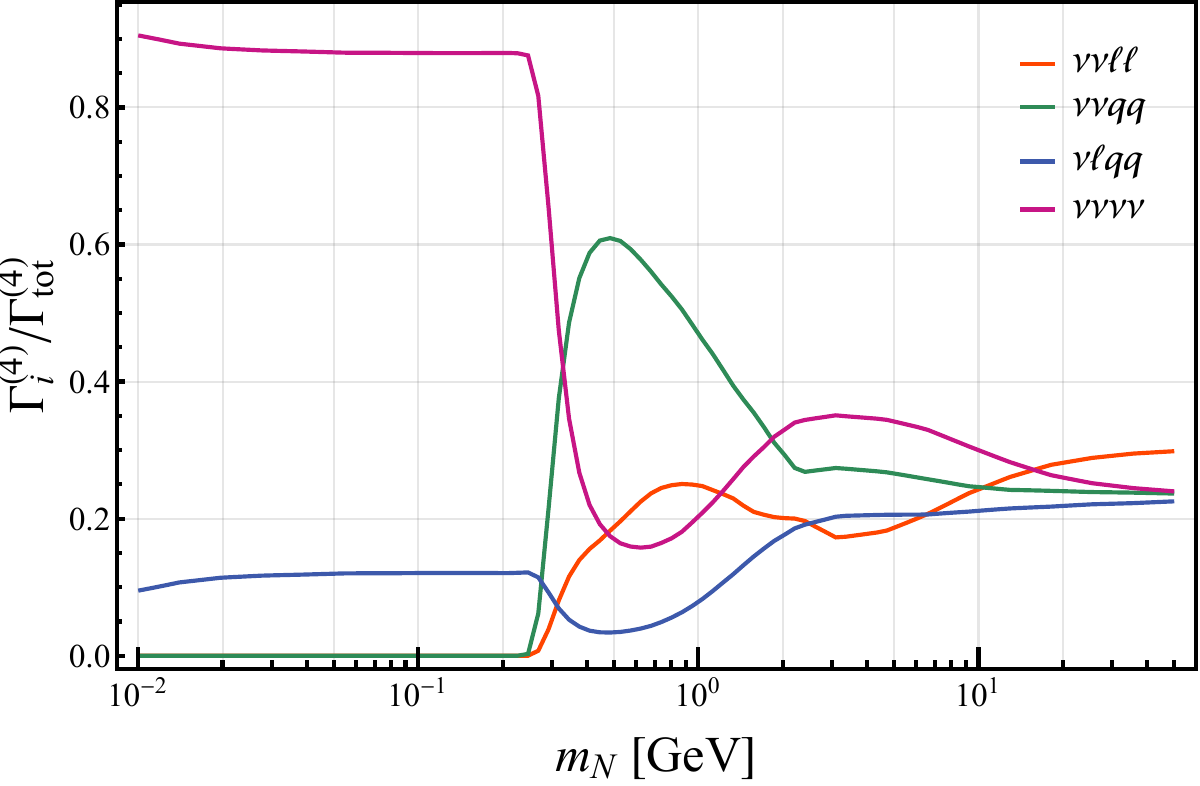}
    \end{subfigure}
    \hfill
    \begin{subfigure}{0.45\textwidth}
        \includegraphics[width=\linewidth]{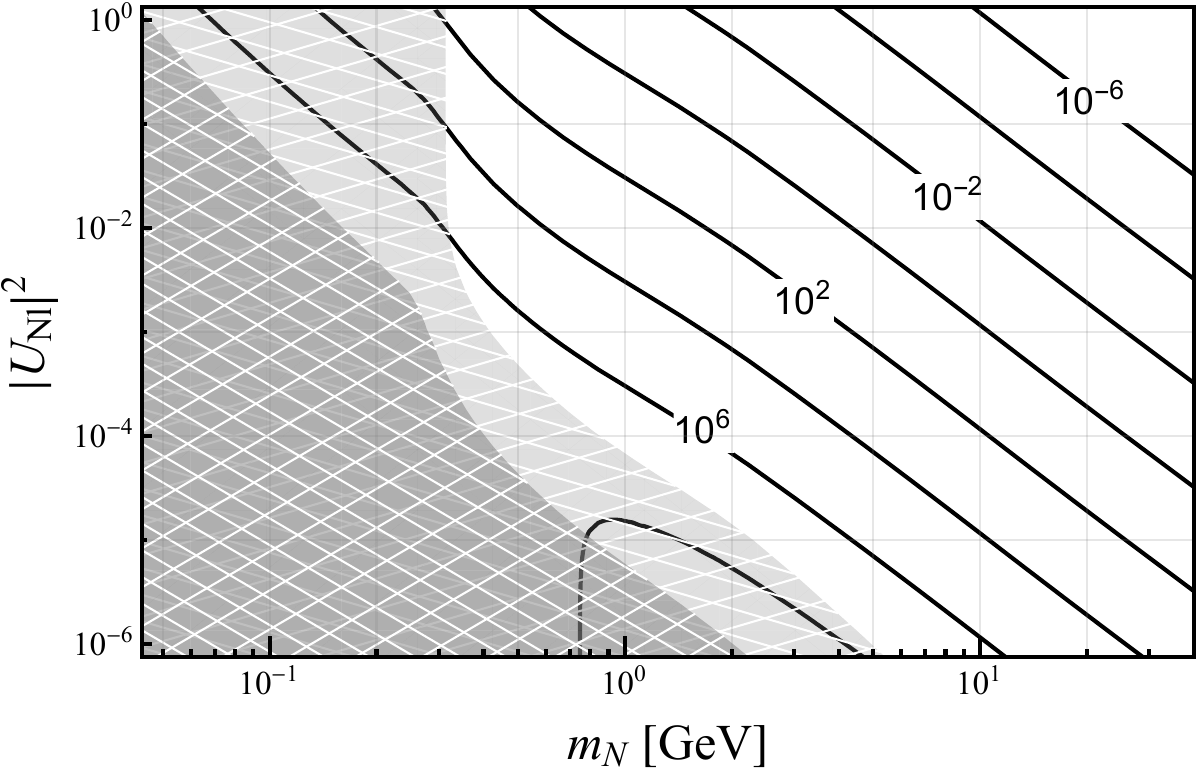}
    \end{subfigure}
    \bigskip
    \caption{Left: Relative decay widths of $\sigma$ to different parton-level four-body final states, as a function of $m_N$. Right: Contours of $c\tau_\sigma$ in meters as a function of $(m_N, |U_{N\ell}|^2)$. Here, we set $\Delta_N=7/4$, $\Delta_{2N}=17/4$ and $m_\sigma/m_N = 0.6$. The light and dark shaded regions correspond to $\mathcal{BR}_\mathrm{inv} > 50\%$ and $\mathcal{BR}_\mathrm{inv} > 99\%$ respectively.}
    \label{fig: sigma_LT}
\end{figure}

Combining all the $\sigma$ decay modes, the leading branching fractions and the contours of $c\tau_\sigma$ are shown in Fig.~\ref{fig: sigma_LT}.

\subsection{Current Constraints}

\label{sec: colliders_beam_dumps}

We now proceed to study the current bounds on this class of models from colliders and beam dumps. We first consider the limits on hidden sector particles produced in the decays of pions, kaons, D- and B-mesons, before turning our attention to production from the decays of $W$ and $Z$ bosons at colliders. 

\subsubsection{Pion and Kaon Decays}

\label{subsec: meson decays}

The most stringent limits on HNLs from pion and kaon decays are based on deviations from the expected kinematic distributions of the visible SM particles produced in the decay. Therefore, these searches are applicable in our case, even if the hidden sector particles decay invisibly or outside the detector. The relevant searches for invisible HNLs are listed in Section~III of \cite{deGouvea:2015euy}. We outline the applicable ones below.

\begin{enumerate}

    \item [(i)] Searches at TRIUMF based on the kinematics of the decay $\pi \rightarrow e + {\rm inv.}$. These exclude $|U_{N\ell}|^2 \gtrsim 10^{-8}$ in the mass range $10~\text{MeV} < m_N < 130~\text{MeV}$~\cite{Britton:1992pg, Britton:1992xv}.

    \item [(ii)] An experiment at KEK based on the kinematics of the decay $K \rightarrow \mu \nu$, which excludes $|U_{N\ell}|^2 \gtrsim 10^{-5}$ in the mass range $70~\text{MeV} < m_N < 300~\text{MeV}$~\cite{Hayano:1982wu}.

    \item [(iii)] The KEK E104 experiment based on the kinematics of the decay $K \rightarrow e \nu$, which excludes $|U_{N\ell}|^2 \gtrsim 10^{-6}$ in the mass range $135~\text{MeV} < m_N < 350~\text{MeV}$~\cite{Yamazaki:1984sj}.

    \item [(iv)] The E949 experiment at BNL based on the kinematics of the decay $K \rightarrow \mu \nu$, which excludes $|U_{N\ell}|^2 \gtrsim 10^{-8}$ in the mass range $175~\text{MeV} < m_N < 300~\text{MeV}$~\cite{E949:2014gsn}.
\end{enumerate}

As is clear from Eqn.~(\ref{eq: meson_two_body_width}), the production rates in our scenario differ from those of a conventional HNL. As a result, we derive the corresponding bounds in our case by rescaling the lowest value of $|U_{N\ell}|^2$ excluded by the above searches by the ratio of the production rates, $\Gamma(X \rightarrow \mathcal{U}_N Y)/\Gamma(X \rightarrow \widetilde{N} Y)$. The resulting constraints from these searches are shown in red in Fig.~\ref{fig: constraints_plot}.

\subsubsection{CHARM} 

The CHARM experiment~\cite{Dorenbosch:164101} searched for decays of long-lived particles produced by colliding the SPS proton beam operating at 400~GeV on a fixed copper target ($N_{PoT} = 1.9\times 10^{18}$ protons on target). The search is sensitive to long-lived particles decaying into a visible $\ell^+ \ell'^-$ pair, where $\ell, \ell' = \{e, \mu\}$. At 90\% CL, scenarios that give rise to a number of signal events, $N_\text{events}>2.3$, are excluded. This search can constrain our model due to composite singlet neutrinos $N$ being produced in the collisions and subsequently decaying to $\sigma \nu$. A signal is obtained if the $\sigma$ particles are sufficiently long-lived to decay to $\ell^+ \ell'^- + {\rm invisible}$ within the detector. 

To determine the bounds on our scenario from this search, we first use \texttt{Pythia~8} to simulate events corresponding to the production of mesons $\mathfrak{m}_i \in \{D^\pm, D^0, D^\pm_s, B^\pm, B^0, B^0_s\}$ at the CHARM experiment. Next, we generate events corresponding to decay $\mathfrak{m}_i \rightarrow \mathcal{U}_N Y_{j}$ in the rest frame of $\mathfrak{m}_i$, by running an accept/reject Monte Carlo over the phase space using the matrix elements given in the appendix. For a given event, depending on the sampled value of $m_\mathcal{U}$, we sample the momenta of the composite states and neutrinos in the final state, following the scheme illustrated in Fig.~\ref{fig: unparticle_to_composites}. We then boost the momentum of $\sigma$ to the lab frame and estimate the efficiency $\epsilon$ of each event. In a coordinate system where the production point is at the origin, the decay region of the CHARM experiment is a cuboid given by $z \in [480~\mathrm{m}, 515~\mathrm{m}]$, $x \in [-1.5~\mathrm{m}, 1.5~\mathrm{m}]$, and $y \in [3.5~\mathrm{m}, 6.5 ~\mathrm{m}]$. For events where $\sigma$ passes through the decay region, we therefore set the efficiency to be,
\begin{equation}
    \epsilon = \mathcal{BR}(\sigma \rightarrow \nu\nu\ell\ell)\times \left[e^{-r_\mathrm{min}/(\gamma c\tau)} - e^{-r_\mathrm{max}/(\gamma c\tau)}\right] \;,
\end{equation}
where $r_\mathrm{min}$ and $ r_\mathrm{max}$ are the distances from the production point at which $\sigma$ enters and exits the decay region, and $\gamma$ is the boost of $\sigma$. For each channel, we average over all the events to obtain the average efficiency, $\overline{\epsilon}$. The number of events is then given by,
\begin{equation}
    N_\mathrm{events} = \sum\limits_i N_\mathrm{prod}(\mathfrak{m}_i)\times \sum\limits_j \mathcal{BR}(\mathfrak{m}_i \rightarrow \mathcal{U}_N Y_j)\overline{\epsilon}^{ij} \;,
\label{eq: CHARM_Nevt}
\end{equation}
where $N_\mathrm{prod}(\mathfrak{m}_i)$ is the number of mesons $\mathfrak{m}_i$ produced at CHARM. The region excluded by CHARM, corresponding to $N_\mathrm{events} > 2.3$ is shown in Fig.~\ref{fig: constraints_plot} (in Brown).

\subsubsection{B-factories}


Both the BaBar and Belle experiments have placed limits on HNLs. The HNL search reported by the Belle collaboration~\cite{Belle:2022tfo} places limits on HNLs produced in $\tau$-decays of the form $\tau^- \rightarrow \pi^- N$ that then decay within the detector volume into a pion and a lepton, $N \rightarrow \pi^\pm \ell^\mp$. In our case, although the $N$ are indeed produced in $\tau$ decays, they promptly decay into a $\sigma$ and a neutrino. Since the $\sigma$ are long-lived and generally decay well outside the detector, the Belle search does not place any significant constraints on our scenario.  

The BaBar experiment also places constraints on HNLs that couple to the $\tau$ lepton~\cite{BaBar:2022cqj}. The search in question is based on kinematic distributions of the visible three-pronged pion system produced in $\tau$ decays, $\tau^\pm \rightarrow 3\pi^\pm + \text{inv}$. The $\tau$ leptons are produced in $e^+e^-$ collisions at $\sqrt{s}=10.58~\text{GeV}$, and their subsequent decays are well separated. In the SM, the invisible final state contains only a single $\nu_\tau$. In the presence of an HNL with $|U_{N\tau}|^2 \neq 0$, the invisible state can also be composed of an $N$, which affects the energy distribution of the $3\pi^\pm$ system. This experimental search does not require that the HNL decay visibly or inside the detector, and as a result, it can be recast to place limits on our scenario. This search results in the strongest limits on couplings to the hidden sector in the mass range $3m_\pi < m_N < m_\tau - 3m_\pi$. The excluded region is shown in Fig.~\ref{fig: constraints_plot} (in green).

\subsubsection{LHC}

Both the ATLAS and CMS collaborations have conducted long-lived particle searches for HNLs. Although the resulting constraints cannot be applied directly, they can be repurposed to place bounds on our scenario. In our case, the final state has two additional neutrinos as compared to the case of conventional HNLs. As a result, we expect the kinematic distributions of the visible final states to deviate somewhat from the distributions considered in these searches. Nevertheless, the results of these searches can be recast to set bounds on our scenario. 

The CMS search of Ref.~\cite{CMS:2023jqi} rules out conventional HNLs in the mass range $2-20~\text{GeV}$ with mixing angles $|U_{N\ell}|^2 \gtrsim 10^{-6}$, based on a final state containing a prompt lepton $\ell_1$, a displaced lepton $\ell_2$, and at least one displaced jet $j$. This search considers events where $\ell_1, \ell_2 \in \{e, \mu\}$. In the case of conventional HNLs, this final state arises from the production of HNLs in $p p \rightarrow W \rightarrow \ell_1 N$ and the subsequent decay of the HNL via $N \rightarrow \ell_2 q q$. In our case, the produced $N$ promptly decays into $\sigma + \nu$, and subsequent decay $\sigma \rightarrow \nu \ell_2 qq$ results in the displaced vertex. The particle flow (PF) algorithm used in this search for event reconstruction has an efficiency of $\epsilon_\mathrm{rec}=95\%$ for transverse displacements of $L_{xy} = 1~\mathrm{cm}$, which gradually drops to zero for transverse displacements of $L_{xy} = 1~\mathrm{m}$. We assume $\epsilon_\mathrm{rec}$ to be a linearly decreasing function of $L_{xy}$ within this range. Furthermore, this search uses the following cuts: $p^{\ell_1}_T > 26~\mathrm{GeV}, ~p^{\ell_2}_T > 3~\mathrm{GeV}, ~p^j_T > 20~\mathrm{GeV}, ~20~\mathrm{GeV} < m_{\ell\ell}< 80~\mathrm{GeV}, ~p^\mathrm{miss}_T < 60~\mathrm{GeV}$, and 70 GeV $< m_{\ell\ell j}< 90~\mathrm{GeV}$. Events passing these cuts are binned into 24 categories corresponding to 4 combinations of the lepton flavors in the final state: $\{ee, e\mu, \mu e, \mu\mu\}$, 3 combinations of the significance of the transverse impact parameter of $\ell_2$ with respect to the primary vertex $d^\mathrm{sig}_{xy}$, and two combinations of the lepton charges: same-sign (SS) and opposite-sign (OS). Given the maximum possible transverse displacement of $L_{xy}=1~\mathrm{m}$ for detection, we find that this search leads to stronger constraints than the limits from precision electroweak measurements in the mass range $m_N \gtrsim 10~\mathrm{GeV}$. As a result, the additional binning of events into boosted and resolved categories (based on $\Delta R_{\ell_2j}$) used in this search is not relevant in our case, and we treat all events as resolved. We proceed to estimate the constraints arising from this search by generating events corresponding to the full chain using \texttt{MadGraph}. We assume perfect detector resolutions and consider the momentum of the jet to be given by the sum of the momenta of the two quarks in the final state. Then, we estimate the efficiency of each event as,
\begin{equation}
    \epsilon^{i} = \epsilon^{i}_\mathrm{cut} \times \int \limits _{s=0}^{s_\mathrm{max}} \frac{ds}{\gamma_i c\tau_\sigma} e^{-s/\gamma_i c\tau_\sigma} \epsilon_\mathrm{rec}(s \sech{\eta_i})\,,
\label{eq: LHC_efficiency}
\end{equation}
where $\epsilon_\mathrm{cut}=1$ if the event passes the cuts, $\gamma_i$ and $\eta_i$ are the boost and pseudorapidity of $\sigma$, and $s_\mathrm{max}$ is the maximum allowed displacement of $\sigma$ such that the corresponding $L_{xy} < 1~\mathrm{m}$. Then, we calculate the average efficiency $\overline{\epsilon}$ over all the events. The total number of expected events over all the bins is given by
\begin{equation}
    N_\mathrm{events} = \mathcal{L}\times \sigma (p p \rightarrow \ell N)\times \mathcal{BR}(\sigma \rightarrow \nu \ell q q)\times \overline{\epsilon} \;,
\label{eq: LHC_N_evt}
\end{equation}
where $\mathcal{L}=139~\mathrm{fb}^{-1}$ is the integrated luminosity and the branching ratio $\mathcal{BR}(\sigma \rightarrow \nu \ell q q)$ is as shown in Fig.~\ref{fig: sigma_LT}. We apply the constraints from this search by excluding the region of the parameter space where $N_\mathrm{events} \geq 1$. This is shown as the light blue region in Fig.~\ref{fig: constraints_plot}.

\begin{figure}[h!]
    \centering
    \begin{subfigure}{0.45\textwidth}
        \includegraphics[width=\linewidth]{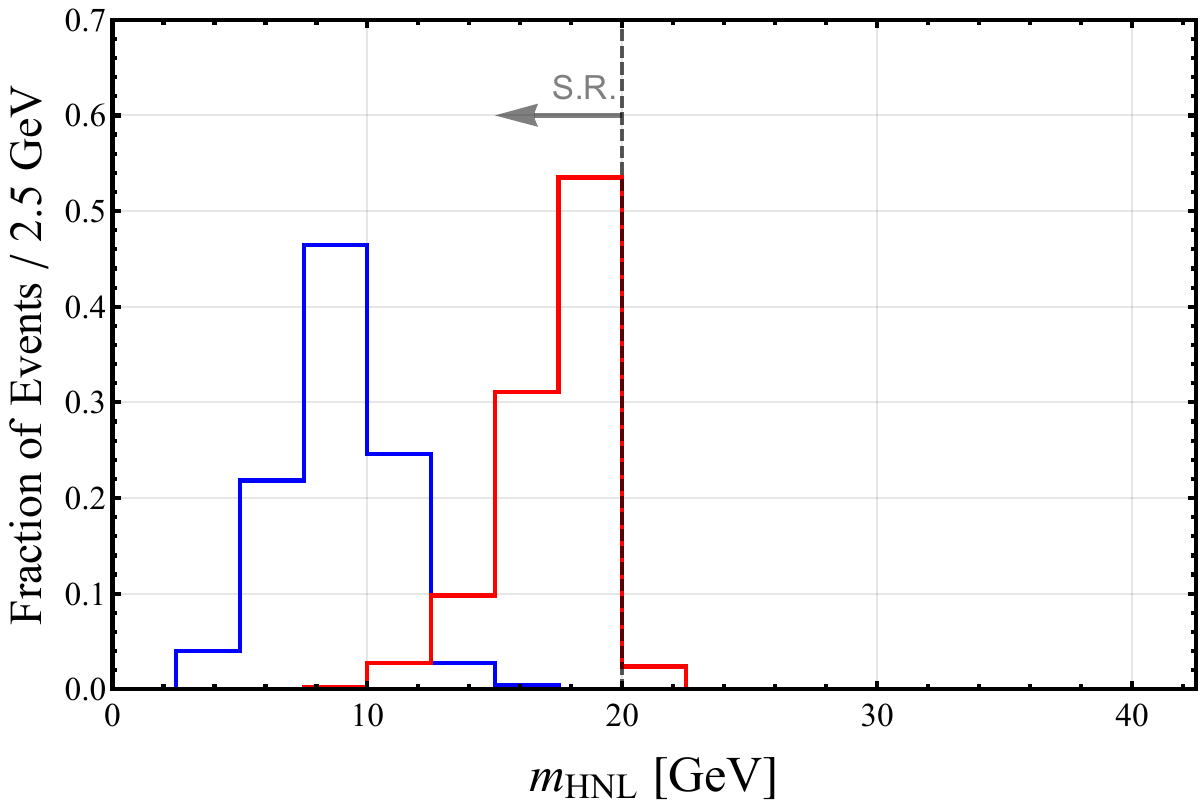}
        \caption{$m_N = 20~\mathrm{GeV}$}
    \end{subfigure}
    \hfill
    \begin{subfigure}{0.45\textwidth}
        \includegraphics[width=\linewidth]{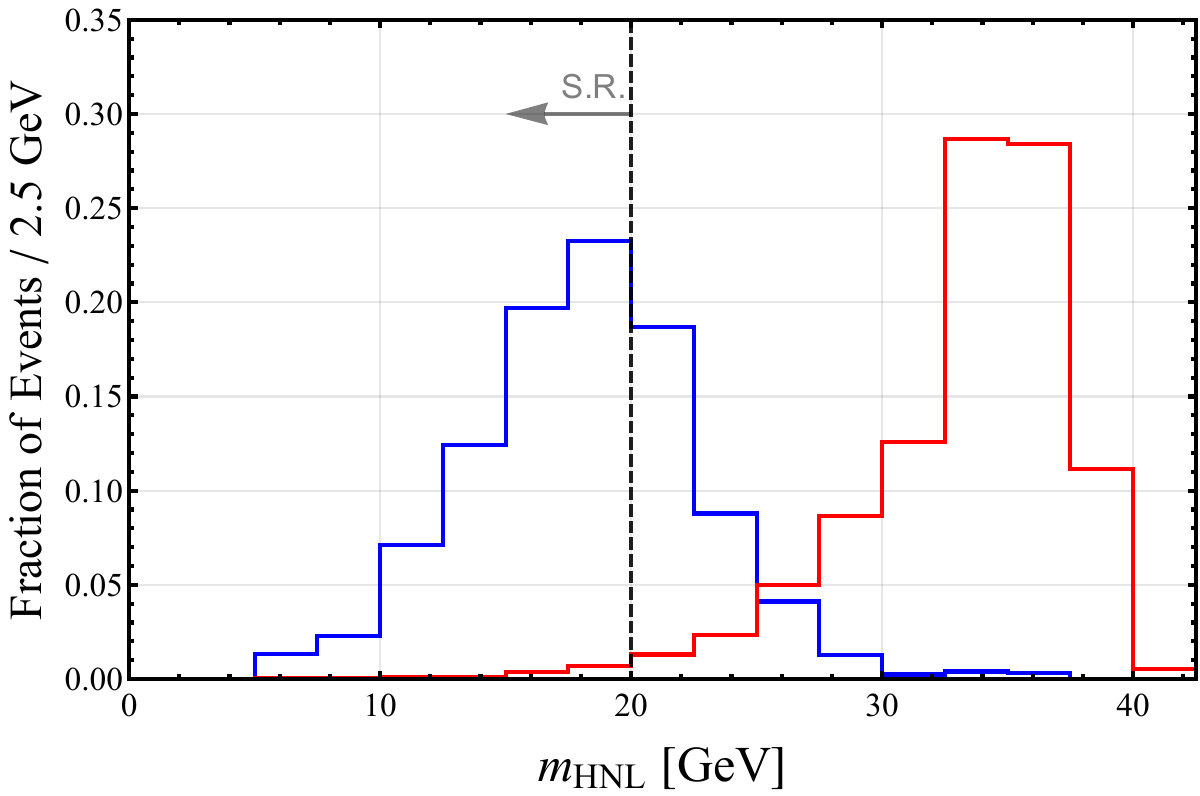}
        \caption{$m_N = 40~\mathrm{GeV}$}
    \end{subfigure}
    \bigskip
    \caption{Distributions of the reconstructed HNL mass $m_{HNL}$ in our case (blue) and in the case of conventional HNLs (red), for events passing all the other kinematic cuts in the ATLAS search of Ref.~\cite{ATLAS:2022atq}. The left and right plots correspond to $m_N = 20~\mathrm{GeV}$ and $m_N = 40~\mathrm{GeV}$ respectively.}
    \label{fig: m_HNL_dist}
\end{figure}

The ATLAS search of Ref.~\cite{ATLAS:2022atq} rules out conventional HNLs in the mass range $2-15~\mathrm{GeV}$ with mixing angles $|U_{N\ell}|^2 \gtrsim 10^{-7}$, based on a final state containing a prompt lepton $\ell_1$ and a displaced vertex (DV) formed by two oppositely charged leptons $\ell_2, \ell_3$. In the case of conventional HNLs, this final state arises from the production of HNLs in $pp \rightarrow W \rightarrow \ell_1 N$ and the subsequent decay of the HNL via $N \rightarrow \nu \ell_2 \ell_3$. In our case, the decay of the dilaton via $\sigma \rightarrow \nu \nu \ell_2 \ell_3$ results in the same visible final state. The cuts used in this search require that at least one lepton has $p_T > 28~\mathrm{GeV}$, $40~\mathrm{GeV} < m_{\rm{DV} + \ell_1} < 90~\mathrm{GeV}$ and that the radial (transverse) position of the DV satisfies $4~\mathrm{mm} < L_{xy} < 300~\mathrm{mm}$. The large-radius tracking (LRT) algorithm employed in this search is efficient for transverse impact parameters up to $d_{xy}=300~\mathrm{mm}$. Then, an algebraic method was used to obtain a reconstructed HNL mass $m_{{\rm HNL}}$ for each event. The signal region was defined by $m_{{\rm HNL}} < 20~\mathrm{GeV}$. No excess of events was found. We proceed to estimate the constraints on our model arising from this search by first generating events corresponding to the full chain using \texttt{MadGraph}. We assume perfect detector resolutions and apply the relevant kinematic cuts on the generated events. Due to the presence of two additional neutrinos in our case, the distributions of $m_{{\rm HNL}}$ are shifted toward lower values as compared to the case of conventional HNLs, as shown in Fig.~\ref{fig: m_HNL_dist}. As a result, the cut on $m_{{\rm HNL}}$ results in higher acceptance rates in our case.

We estimate the efficiency of each event in a manner analogous to Eqn.~(\ref{eq: LHC_efficiency}), where $\epsilon_\mathrm{rec}$ is instead a linearly decreasing function of $d_{xy}$. Using the average efficiency $\overline{\epsilon}$, we estimate the total number of expected events as in Eqn.~(\ref{eq: LHC_N_evt}), but where the branching ratio is replaced by $\mathcal{BR}(\sigma \rightarrow \nu \nu \ell \ell)$. The region of the parameter space where $N_\mathrm{evt} \geq 1$ is excluded is shown in Fig.~\ref{fig: constraints_plot}.

\subsection{Ongoing and Future Searches}

\subsubsection{FASER 2}
We now turn our attention to ongoing and future experiments, beginning with the proposed FASER 2 detector. FASER 2 searches for light, long-lived particles that are produced in the extreme forward direction in proton-proton collisions at the ATLAS experiment. The FASER 2 detector is to be placed in the forward region of the ATLAS interaction point (IP), at a distance of $L_\text{max} = 480{\rm m}$. We employ the detector specifications presented in Ref.~\cite{Kling:2018wct}, where the authors estimated the reach of FASER~2 for the case of conventional HNLs. These are given by:
\begin{equation*}
    \begin{split}
        \Delta &= 10~\text{m},\\
        R &= 1~\text{m},\\
        \mathcal{L} &= 3~\text{ab}^{-1},
    \end{split}
\end{equation*}
where $\Delta$ is the depth and $R$ is the radius of the cylindrical detector. With these dimensions, the detector is only sensitive to dilatons $\sigma$ that are produced at an angle $\theta_\sigma < \theta_\text{max} = 2.1\times 10^{-3}$ with respect to the positive beam axis. This restriction on $\theta_\sigma$ leads to a reasonable acceptance for $\sigma$ particles produced in heavy meson decays, but it greatly suppresses the reach for $\sigma$ particles from Drell-Yan production, which have a more central angular distribution. We have confirmed these expectations using Monte Carlo simulations. 

\begin{figure}[h!]
    \centering
    \begin{subfigure}{0.45\textwidth}
        \includegraphics[width=\linewidth]{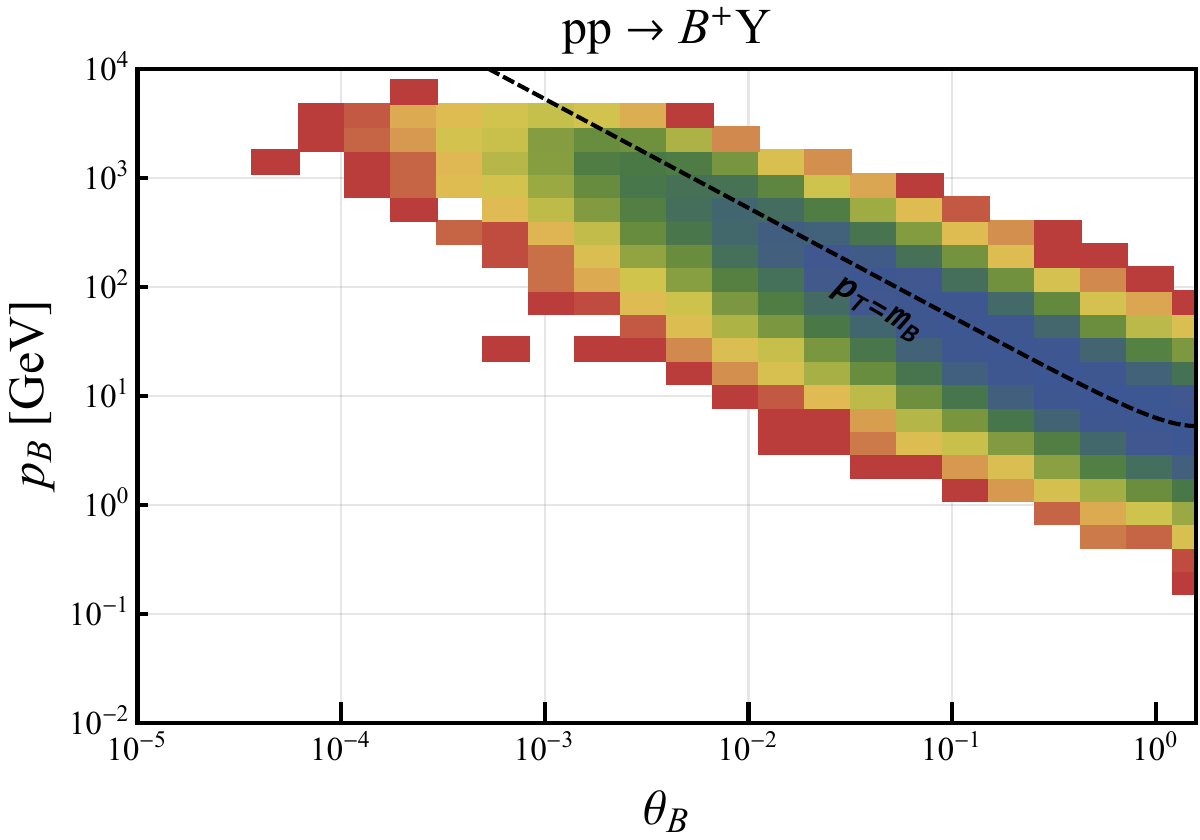}
    \end{subfigure}
    \hfill
    \begin{subfigure}{0.45\textwidth}
        \includegraphics[width=\linewidth]{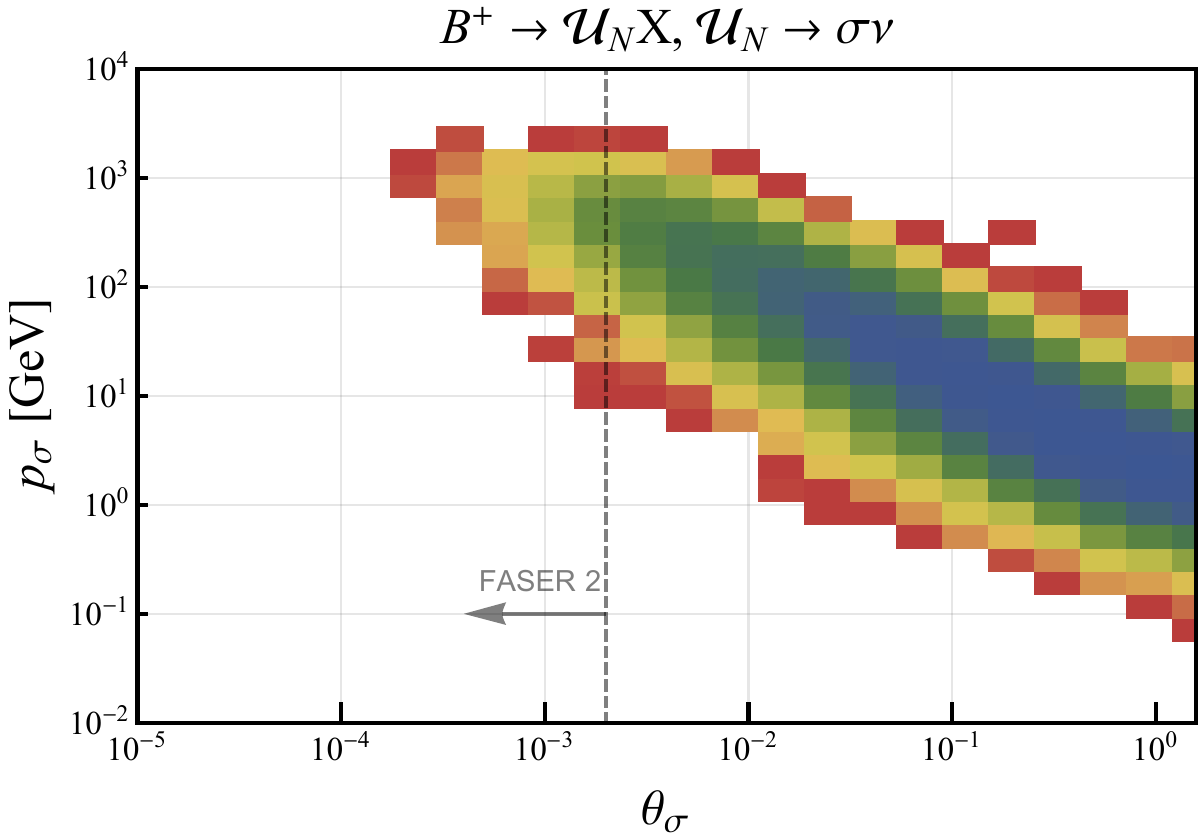}
    \end{subfigure}
    \bigskip
    \caption{(Left) Distributions in the $(p, \theta)$ plane for $B^+$ mesons produced in proton-proton collisions at $\sqrt{s}=14~\mathrm{TeV}$, where the events are generated using \texttt{Pythia 8}. (Right) Similar distributions corresponding to $\sigma$ produced in the decays of $B^+$ mesons, where $m_N = 2~\mathrm{GeV}$. The dashed vertical line corresponds to $\theta=\theta_\mathrm{max}$.}
    \label{fig: FASER_distributions}
\end{figure}

Since the detector efficiencies for FASER 2 have not yet been studied, we will assume a $100\%$ signal efficiency for the final states that contain a visible pair of leptons $\ell^+_1 \ell^-_2$, where $\ell_1, \ell_2 \in \{e, \mu\}$. We limit our analysis to $\{e, \mu\}$, due to the inefficiencies involved in the identification and reconstruction of $\tau$-leptons. With these requirements, we calculate an upper bound on the number of expected signal events at FASER~2 following a methodology similar to that outlined in Eqn.~(\ref{eq: CHARM_Nevt}). In order to estimate the geometric efficiencies $(\overline{\epsilon}^{ij}_\mathrm{det})$, we generate events corresponding to the production of pseudoscalar mesons $\mathfrak{m}_i$ in proton-proton collisions at $\sqrt{s} = 14~\mathrm{TeV}$ using \texttt{Pythia~8}. For comparison with Ref.~\cite{Kling:2018wct}, where events were generated for conventional HNLs (using FONLL for the heavy meson distributions), two-dimensional distributions in the $(p, \theta)$ plane for $B^+$ mesons and for dilatons $\sigma$ produced from decays of $B^+$ mesons are shown in Fig.~\ref{fig: FASER_distributions}. 

For each meson $\mathfrak{m}_i$, we sample events until a sufficient number of events $(> 1500)$ has been generated within the narrow angular acceptance range of FASER~2. This allows us to get reasonable estimates for the average geometric acceptances, and for the distributions of the boost factors of the $\sigma$ entering the decay volume. For the numbers of mesons produced $N_\mathrm{prod}(\mathfrak{m}_i)$, we use the values reported in Ref.~\cite{Kling:2018wct}. The contour for $N_\text{events} = 3$ is shown in Fig.~\ref{fig: constraints_plot} (in blue).

\subsubsection{SHiP}

The proposed Search for Hidden Particles (SHiP) experiment~\cite{SHiP:2018yqc} will search for long-lived particles produced in the collisions of a 400~GeV proton beam with a fixed molybdenum target ($N_{PoT}=2\times 10^{20}$ protons on target). The decay volume of SHiP will be a 50~m long pyramidal frustum with upstream and downstream dimensions of $1.5~\mathrm{m}\times 4.3~\mathrm{m}$ and $5~\mathrm{m}\times 10~\mathrm{m}$ respectively. The upstream end of the decay region will be located at a distance of $L_\mathrm{min}=45~\mathrm{m}$. Calorimeters will be placed at the downstream end to detect the visible SM particles produced in the decays of the long-lived particle. As a result, this experiment is sensitive to dilatons $\sigma$ produced in the angular region $\theta < \theta_\mathrm{max} = 9.5\times 10^{-2}$ that decay into various visible final states. Due to the small angular region of acceptance, the $\sigma$ that enter the decay volume are predominantly produced from the decays of heavy mesons. 

To estimate the reach of SHiP, we first generate events corresponding to the production of D and B mesons using \texttt{Pythia 8}. We consider an event to be accepted if $\sigma$ is produced with $\theta < \theta_\mathrm{max}$ and decays within the detector. We use the values of $N_\mathrm{prod}(\mathfrak{m}_i)$ reported in Ref.~\cite{SHiP:2018xqw}, where the reach of SHiP has been estimated for the case of conventional HNLs. Additionally, we consider a $100\%$ detection efficiency for the channels $\sigma \rightarrow \nu\ell q q$ and $\sigma \rightarrow \nu\nu\ell\ell$, since these channels result in at least two charged tracks. The contour for $N_\text{events} = 3$ is shown in Fig.~\ref{fig: constraints_plot} (in purple).

\subsubsection{Belle~II}

\begin{figure}[h!]
    \centering
    \includegraphics[width=0.7\textwidth]{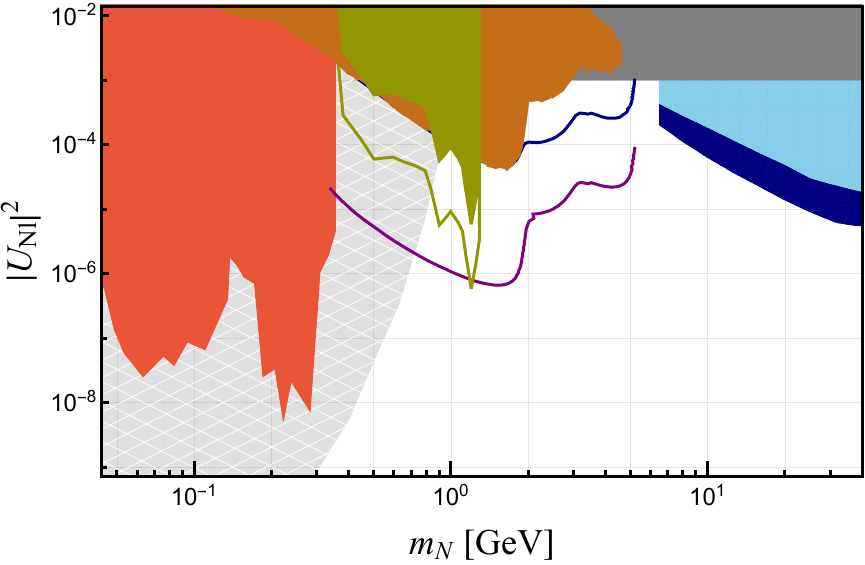}
    \caption{We show the exclusion (shaded) and sensitivity (curves without shading) regions for collider and beam dump experiments. The region shown in brown is excluded by CHARM. The BaBar search for invisible HNLs excludes the region shown in green. The regions shown in dark blue and light blue are excluded by the ATLAS and CMS searches respectively. The region shaded in red is excluded by invisible searches based on kaon and pion decays. In the mesh shaded region below a GeV, constraints from supernova dynamics may potentially apply (see Section~\ref{sec: astro}). The green, blue and purple curves show the projected reach for Belle~II, FASER~2 and SHiP respectively.}
    \label{fig: constraints_plot}
\end{figure}

The Belle~II experiment~\cite{Belle-II:2010dht} is expected to reach a total integrated luminosity of $50~{\rm ab}^{-1}$ at $\sqrt{s} = 11~{\rm GeV}$. This is about $100$ times more than that of the BaBar experiment. As a result, $\sim 100$ times as many $\tau^\pm$ are expected to be produced at Belle~II. Therefore, we expect that Belle~II will have much greater reach than BaBar in the $(\tau^\pm \rightarrow 3\pi^\pm + \text{inv.})$ channel that was studied at Babar. To estimate the reach, we assume that the detector resolution at Belle~II will be comparable to that of BaBar and that the search will be statistics-dominated.  Under these assumptions, Belle~II is expected to probe additional parameter space where the minimum number of signal events (which scales as $|U_{N\ell}|^2$) is $\sim 0.1$ times that of BaBar. Therefore, the reach of Belle~II can be estimated by scaling the BaBar curve by this factor. The is shown in Fig.~\ref{fig: constraints_plot} as the green contour.

\subsection{Discussion}

\begin{figure}[h!]
    \centering
    \begin{subfigure}{0.45\textwidth}
        \includegraphics[width=\linewidth]{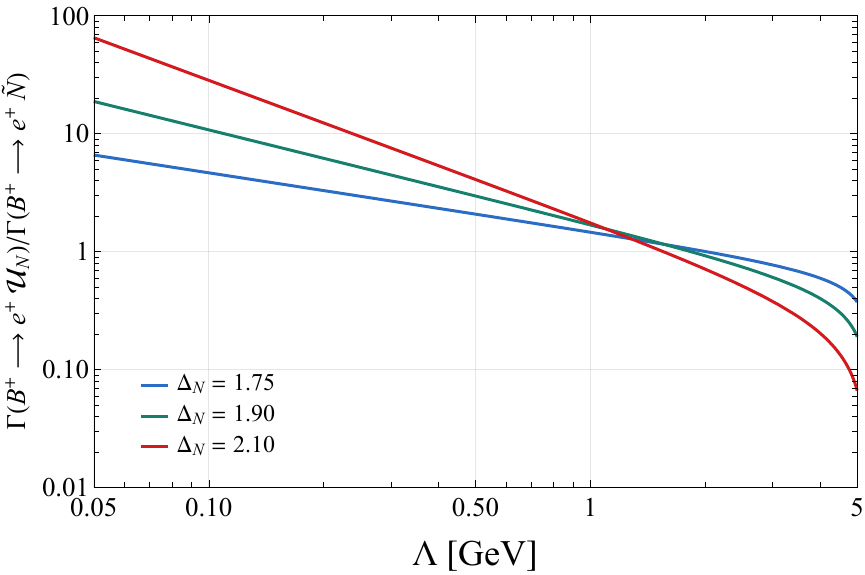}
    \end{subfigure}
    \hfill
    \begin{subfigure}{0.45\textwidth}
        \includegraphics[width=\linewidth]{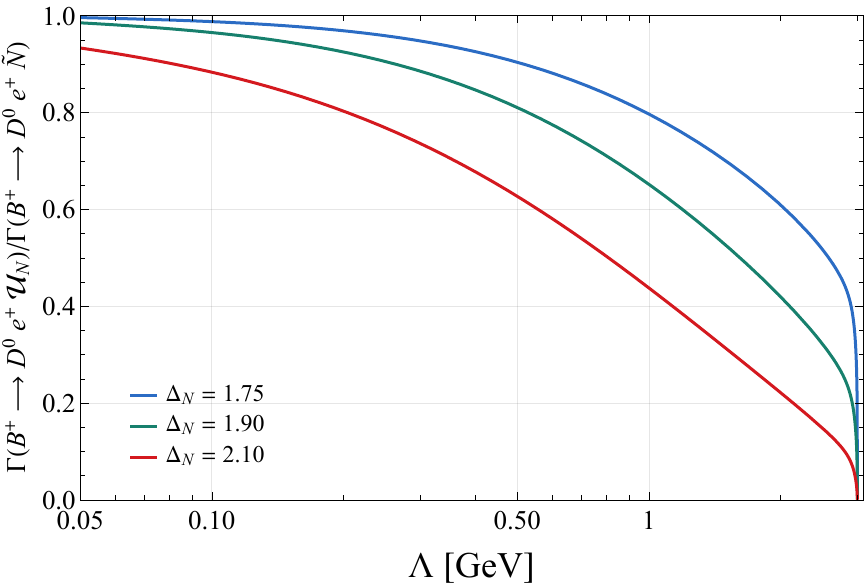}
    \end{subfigure}
    \bigskip
    \caption{A comparison of the partial decay widths of two-body (left) and three-body (right) decays of the $B^+$ meson containing $\mathcal{U}_N$ in the final state for different $\Delta_N$, with respect to the case of a conventional HNL.}
    \label{fig: prod_rates_meson}
\end{figure}

We see from Fig.~\ref{fig: constraints_plot} that FASER, SHiP and Belle II all have the potential to explore new parameter space. We now discuss how these results depend on the scaling dimension $\Delta_N$ of the operator $\mathcal{O}_N$. For values of $m_N$ below 350 MeV, the strongest limits are from the two-body decays of light mesons into a charged lepton and invisible hidden sector states. For a fixed value of $|U_{N \ell}|^2$, the branching ratio for this process depends sensitively on the available phase space and also on the value of $\Delta_N$. As can be seen Fig.~\ref{fig: prod_rates_meson}, the partial decay widths of such decays are enhanced for low values of $\Lambda$ and larger $\Delta_N$ as compared to the case of conventional HNLs. However, for $\Lambda \gtrsim 0.3 \times (m_\mathfrak{m} - m_\ell)$, we see that the partial decay widths are comparatively reduced, with greater suppression for larger $\Delta_N$. Due to this, the constraints from the various searches discussed in Section~\ref{subsec: meson decays} may be stronger or weaker than for a conventional HNL, depending in detail on the precise values of $m_N$ and $\Delta_N$.

\begin{figure}[h!]
    \centering
    \includegraphics[width=0.45\linewidth]{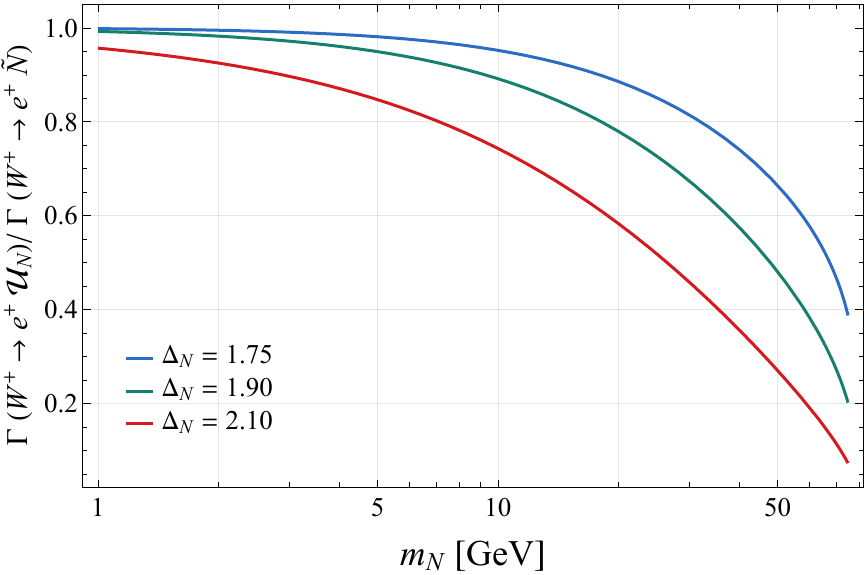}
    \caption{A comparison of the partial decay widths of $W \rightarrow \ell \mathcal{U}_N$ for different $\Delta_N$, with respect to the case of a conventional HNL.}
    \label{fig: prod_rates_DY}
\end{figure}

\begin{figure}[h!]
    \centering
    \begin{subfigure}{0.45\textwidth}
        \includegraphics[width=\linewidth]{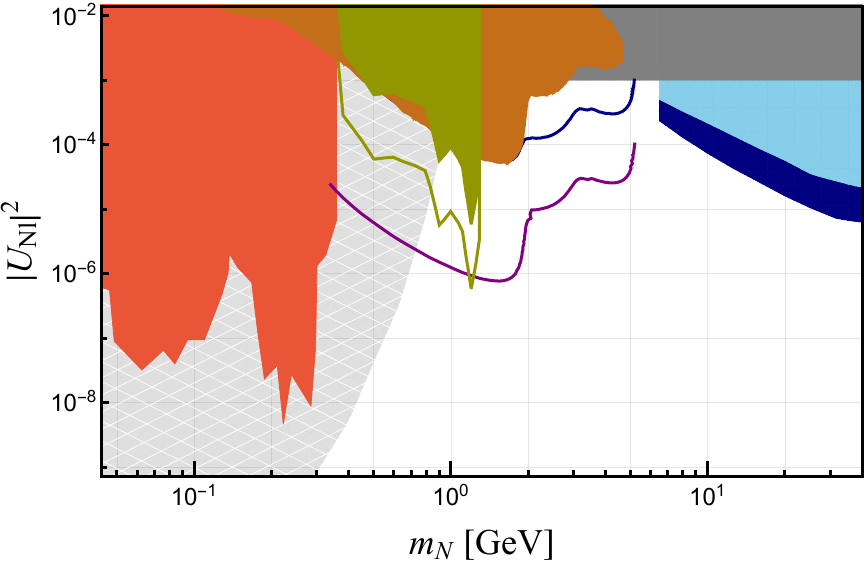}
    \end{subfigure}
    \hfill
    \begin{subfigure}{0.45\textwidth}
        \includegraphics[width=\linewidth]{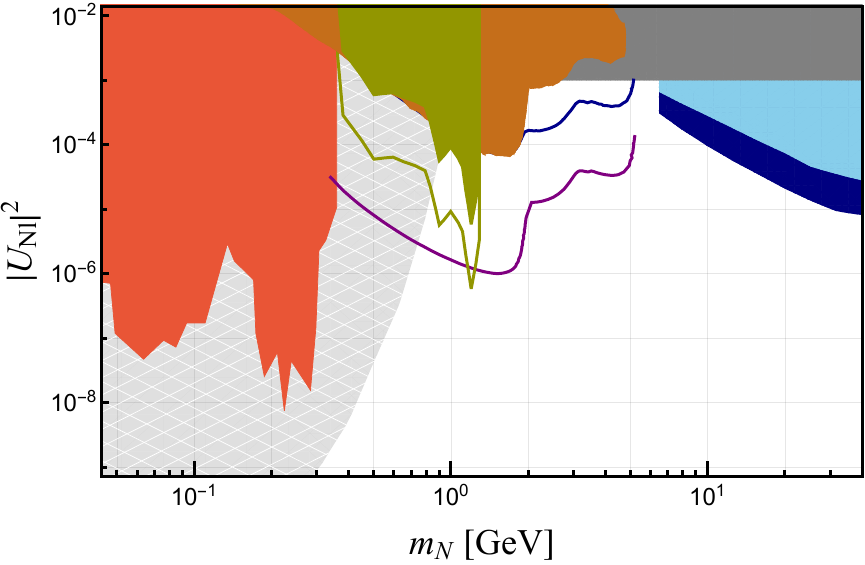}
    \end{subfigure}
    \bigskip
    \caption{Same as Fig.~\ref{fig: constraints_plot}, but for $\Delta_N = 1.9$ (left) and $\Delta_N = 2.1$ (right). The constraints from astrophysics (gray) are largely unchanged. The constraints from BaBar and the estimated reach of Belle~II also remain largely unchanged, since they are independent of the dilaton lifetime. However, the constraints from CMS, ATLAS and CHARM, as well as the projected reach of FASER~2 and SHiP, are strongly dependent on $c\tau_\sigma$, which is in turn dependent on $\Delta_N$.}
    \label{fig: constraints_plot_Delta_Ns}
\end{figure}

For values of $m_N$ above 350 GeV, searches based on the decays of charmed and B-hadrons come into play. In this case, the constraints and the expected reach also depend on $c\tau_\sigma$ and geometric efficiencies. Since $c\tau_\sigma$ is inversely proportional to $(2\Delta_N - 5)^2$ in the relevant parameter space, for fixed $|U_{N\ell}|^2$ and $m_N$, higher values of $\Delta_N$ generally result in reduced sensitivity. The Babar search based on 3-prong tau decays does not require that the hidden sector decay visibly, and so the bound is less dependent on the value of $\Delta_N$. This also holds for the proposed search at Belle-II.
In the case of ATLAS and CMS, the production rates of $\mathcal{U}_N$ through Drell-Yan are suppressed at higher values of $\Delta_N$, as shown in Fig.~\ref{fig: prod_rates_DY}, resulting in weaker constraints. For comparison, we show the exclusion and sensitivity regions for $\Delta_N = 1.9$ and $\Delta_N = 2.1$ in Fig.~\ref{fig: constraints_plot_Delta_Ns}.


\section{Cosmology}
\label{sec: cosmo}

Next, we turn our attention to the cosmological history of this class of models. As shown in Ref.~\cite{Chacko:2020zze}, the lower bound on the mixing angle in Eq.~(\ref{U_Nl_range}) ensures that the hidden sector is in equilibrium with the SM neutrinos at temperatures of order the compositeness scale $\Lambda$ through the processes $\nu\nu\rightarrow \mathcal{U}_N$ and $\nu\,\mathcal{U}_N \rightarrow \mathcal{U}_N$. Once the temperature falls below the compositeness scale, the hidden sector states begin to exit the thermal bath, and will eventually decouple from the SM. We wish to determine the temperature $T_{\rm dec}$ at which this happens, and to thereby establish that the decoupling occurs too early to affect Big Bang nucleosynthesis (BBN) or the cosmic microwave background (CMB).

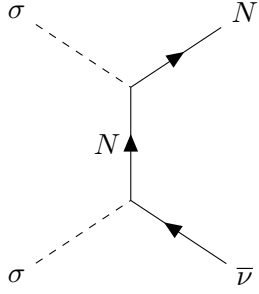
\begin{figure}[h!]
\centering
\begin{tikzpicture}[scale=0.5]
    \begin{feynman}
        \vertex (i1) at (-3, 3.5) {\(\sigma\)};
        \vertex (i2) at (-3, -3.5) {\(\sigma\)};
        \vertex (a) at (0, 1.5);
        \vertex (b) at (0, -1.5);
        \vertex (f1) at (3, 3.5) {\(N\)};
        \vertex (f2) at (3, -3.5) {\(\overline{\nu}\)};
        \diagram*{
        (i1) -- [scalar] (a),
        (i2) -- [scalar] (b),
        (b) --  [fermion, edge label ={\(N\)}] (a),
        (a) -- [fermion] (f1),
        (b) -- [anti fermion] (f2),
        };
    \end{feynman}
\end{tikzpicture}
\caption{Feynman diagram for the leading process keeping the hidden sector in thermal equilibrium.}
\label{fig:feynman-eq}
\end{figure}

At temperatures below $\Lambda$, the hidden sector states begin going out of the bath, with the heavier states annihilating or decaying into lighter ones. However,
since the dilaton $\sigma$ is the lightest state in the hidden sector, it can only annihilate into final states containing a SM particle. The dominant annihilation process for $\sigma$ is $\sigma \sigma \rightarrow \bar{\nu} N$, since this process involves just a single factor of the mixing angle $U_{Nl}$ at the amplitude level (see Fig.~\ref{fig:feynman-eq}). This process is kinematically allowed since $2m_\sigma > m_N$ for $m_\sigma/m_N = 0.6$. The $N$ in the final state subsequently decays back to a dilaton and a neutrino.

\begin{figure}[h!]
    \centering
    \includegraphics[width=0.7\textwidth]{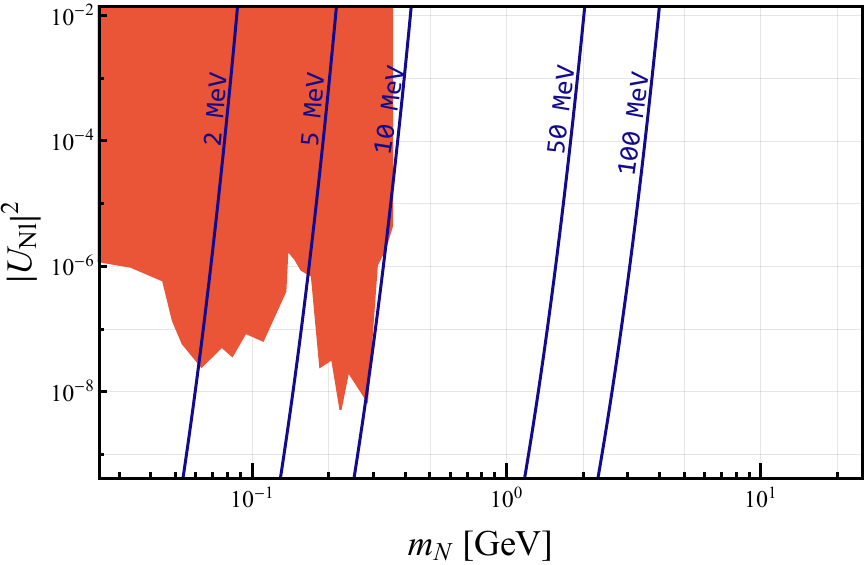}
    \caption{Contours of the decoupling temperature $T_{dec}$ as a function of $(m_N, |U_{Nl}|^2)$. The shaded region in red is excluded by light meson decays (Section~\ref{subsec: meson decays}).}
    \label{fig: bbn1}
\end{figure}

The cross section for the annihilation process scales as
\begin{equation}
\begin{split}
    \sigma \sim (2\Delta_N - 5)^2 \pi^3 |U_{N\ell}|^2 \frac{m_N}{s \sqrt{s - 4m^2_\sigma}} \;,
\end{split}
\end{equation}
from which we can estimate the interaction rate as
\begin{equation}
    n_\sigma \langle \sigma v \rangle \sim  |U_{Nl}|^2 T \sqrt{\frac{T}{m_\sigma}}e^{-{m_\sigma}/{T}} \;.
\end{equation}
 The exponential suppression in this expression is because the number density of $\sigma$ particles falls away rapidly at temperatures below their mass. The hidden sector decouples when $n_\sigma \langle \sigma v \rangle \sim H$, where the Hubble rate $H \sim T^2/M_{Pl}$. From this, we can estimate the decoupling temperature $T_{\rm dec}$ by solving for 
 \begin{equation}
    |U_{Nl}|^{2} \sqrt{\frac{M^2_{Pl}}{T_{dec} m_\sigma}} e^{-{m_\sigma}/{T_{\rm dec}}} = 1 \,,
\label{eq: bbn1}
 \end{equation}
 where we have neglected the $\mathcal{O}(1)$ factors. Contours of $T_{\rm dec}$ are shown in Fig.~\ref{fig: bbn1}. 

As can be seen from the plot, for compositeness scales $\Lambda \gtrsim$ 100 MeV, the hidden sector decouples from the SM at temperatures $T_{\rm dec} > 10~\text{MeV} \sim 0.01~\text{s}$. Therefore, decoupling occurs prior to the BBN epoch, which begins around $t_{\rm BBN} \sim 1~\text{s}$. Moreover, as can be seen from Fig.~\ref{fig: sigma_LT}, the dilaton lifetime $\tau_\sigma$ is less than 1 s in the entire parameter space of interest. Hence, most of the relic $\sigma$ particles decay back into the SM well before BBN. As a result, early universe cosmology at temperatures below an MeV is identical to that in the SM, and so this class of models is consistent with the predictions from BBN and observations of the CMB.

\section{Astrophysical Constraints}
\label{sec: astro}

We now turn our attention to the constraints on this class of models from astrophysical considerations. Since our focus is on compositeness scales of the order of 100 MeV or higher, which is well above the temperature in stars, only the constraints arising from supernovae need to be considered. If produced in abundance, the emission of hidden sector particles from the supernova core can lead to excessive cooling, preventing the explosion from taking place. Even in the case that the hidden sector particles that are produced remain trapped in the core, they may affect the supernova dynamics if their contribution to the total energy density is significant. In this section, we calculate the rate of energy transfer to the hidden sector and determine the range of parameters for which the supernova dynamics is potentially affected. We base the calculation on a set of typical parameters for Type-II supernovae: core temperature $T_{\rm SN} = 30~\text{MeV}$, supernova lifetime $t_{\rm SN} = 10~{\rm s}$, and supernova radius $L_{\rm SN} = 10~\text{\rm km}$. We take the baryon number density inside the core as $n_B = 0.2~\text{fm}^{-3} \approx 1.2\times 10^{-3}~\text{GeV}^3$ and the proton fraction as $Y_p = 0.3$. Then the chemical potentials of the protons and the neutrons are given by $\mu_p = -55~\text{MeV}$, $\mu_n = -8~ \text{MeV}$ \cite{Constantinou:2014hha}. Since the core is electrically neutral and the processes $p^+ + e^- \leftrightarrow n + \nu_e$ are in chemical equilibrium, the chemical potentials of the electrons and neutrinos are then given by $\mu_e = 207~\text{MeV}, ~\mu_\nu = 160~\text{MeV}$. 

We begin by calculating the rate of energy transfer to the hidden sector. Taking the chemical potentials into account, we find that the production of hidden sector states through channels that involve nucleons in the initial state (such as $e^- p^+ \rightarrow N n$, etc.) is suppressed as compared to the rates from channels that involve only neutrinos (and antineutrinos) in the initial state. The two most dominant channels are, 
 \begin{equation}
 \begin{split}
\sigma(\nu \bar{\nu} \rightarrow \sigma \sigma) &\approx \frac{16\pi^3 (2\Delta_N - 5)^4 |U_{N\ell}|^4}{30m^2_\sigma}\left(\frac{m^2_\sigma}{m^2_\sigma + m^2_N}\right)^4 \left(1 - \frac{4m^2_\sigma}{s}\right)^{5/2} \;, \\
\sigma (\nu \nu \rightarrow N \nu) &\approx \frac{\pi^3 |U_{N\ell}|^6}{m^2_N}\left(1 - \frac{m^2_N}{s}\right)^2 \; .
\end{split}
\label{eq: prod_cs_SN}
\end{equation}
The cross section for $\nu \nu \rightarrow N \nu$ above is dominated by dilaton exchange. Although the corresponding amplitude is suppressed by three powers of the mixing angle, this is larger than the contribution from the weak interactions. This differs from a conventional HNL, for which the dilaton exchange contribution is absent. The cross sections given in Eqn.~(\ref{eq: prod_cs_SN}) represent the leading order terms in the expansions around the threshold values of $s_0 = 4m^2_\sigma$ and $s_0 = m^2_N$, respectively.

The energy transferred into the hidden sector over the lifetime of the supernova from each of these processes can then be estimated as
  \begin{equation}
    \begin{split}
        E_{\rm HS} (\nu \bar{\nu} \rightarrow \sigma \sigma) &= 2 n_\nu n_{\bar{\nu}} m_\sigma \langle \sigma v \rangle V_{\rm SN} t_{\rm SN}\\
                   &\approx \frac{128\pi (2\Delta_N - 5)^4 |U_{N\ell}|^4 m_\sigma}{3}  \left(\frac{T_\mathrm{SN}}{2m_\sigma}\right)^5 \left(L^3_\mathrm{SN}t_\mathrm{SN} m^4_\sigma\right) e^{-{2m_\sigma}/{T_{\rm SN}}} \;, \\
        E_{\rm HS} (\nu \nu \rightarrow N \nu) &=  n^2_\nu m_N \langle \sigma v \rangle V_{\rm SN} t_{\rm SN} \\ 
                    &\approx \frac{2|U_{N\ell}|^6 m_N}{3}\sqrt{\frac{\pi}{2}}\left(\frac{T_\mathrm{SN}}{m_N}\right)^{9/2} \left(L^3_{\rm SN} t_{\rm SN} m^4_N\right) e^{-{(m_N - 2\mu_\nu)}/{T_{\rm SN}}} \;.
    \end{split}
    \label{eq: SN_prod}
 \end{equation}

\begin{figure}[h!]
     \centering
     \includegraphics[width=0.7\textwidth]{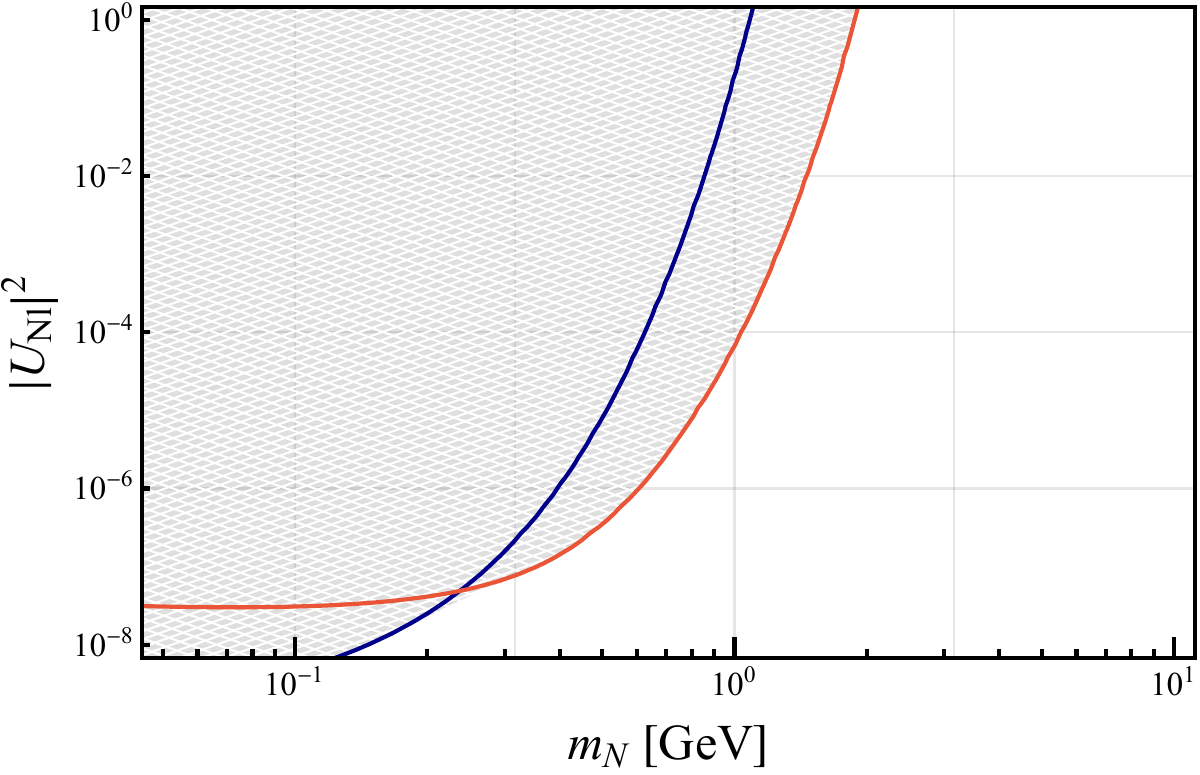}
     \caption{Contours of $E_{\rm HS}/ E_{\rm SN} = 100\%$ for production through $\nu \bar{\nu} \rightarrow \sigma \sigma$ (blue) and $\nu \nu \rightarrow N \nu$ (red). The shaded region represents the parameter space where the total energy transferred into the hidden sector is larger than $\mathcal{F}=100\%$ of $E_{\rm SN} \sim 10^{56}~{\rm GeV}$.}
     \label{fig: sn_prod}
 \end{figure}

  Here $V_{\rm SN} = (4\pi/3) L_{\rm SN}^3$ represents the volume of the supernova core. Which of these two processes dominates depends on the values of $m_N$ and $|U_{N\ell}|^2$. The total energy output of the supernova is of the order of $E_{\rm SN} \sim 10^{53}~\text{ergs} \sim 10^{56}~\text{GeV}$, much of which is carried away by the ejected neutrinos. If $E_{\rm HS}$, the total energy transferred into the hidden sector, is comparable to or larger than this, the supernova luminosity can potentially be affected. Contours for $E_{HS}/E_{SN} \equiv \mathcal{F} = 100\%$ are shown in Fig.~\ref{fig: sn_prod}, where we used the full expressions for the cross-sections and evaluated $\langle \sigma v \rangle$ using numerical integration. Next, we outline two possible constraints in the region where $\mathcal{F}$ is sizable.

\begin{figure}[h!]
    \centering
    \includegraphics[width=0.7\textwidth]{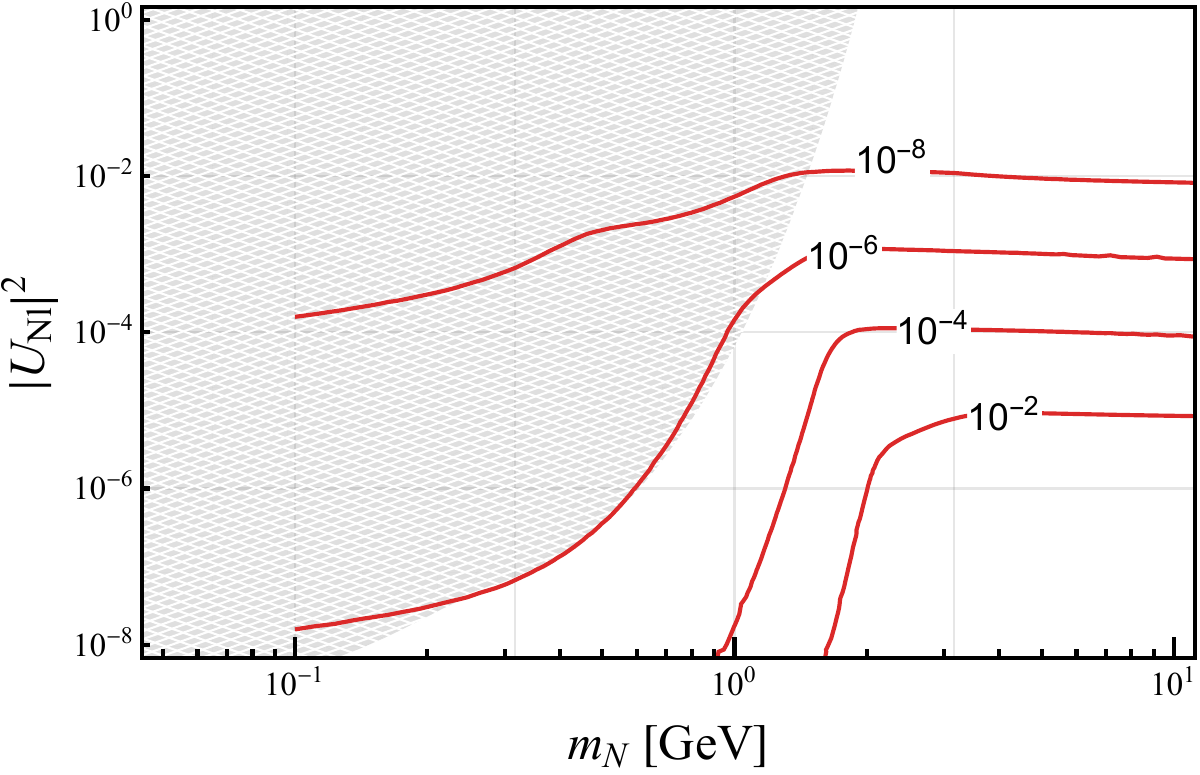}
    \caption{Contours of $\bar{d}/L_{\rm SN}$, as a function of $(m_N, |U_{N\ell}|^2)$. The shaded region represents the parameter space for which $E_{HS}/E_{SM} > 100\%$. As can be seen here, $\bar{d} \ll L_{\rm SN}$ in the region where any significant amount of energy is transferred into the hidden sector. As a result, the hidden sector particles remain trapped inside the supernova core.}
    \label{fig: sn_esc}
\end{figure}

If, once produced, the hidden sector particles are able to escape the core, the observed luminosity of the supernova will be reduced~\cite{Raffelt:1996wa}. Any composite singlet neutrinos that are produced will quickly decay into a dilaton and a neutrino. Therefore, the condition for this to happen is that the typical distance $\bar{d}$ traveled by a dilaton $\sigma$ is larger than the size of the supernova, $(\bar{d} > L_{SN})$. In our case, $\bar{d}$ is set by the elastic scattering process $\sigma \nu \rightarrow \sigma \nu$. For this process, the scattering cross section is given by
 \begin{equation}
    \sigma_\text{el} = \frac{(2\pi)^4 (2\Delta_N - 5)^4 |U_{N\ell}|^4}{16 \pi s} \left[\frac{(s - m^2_\sigma)^2}{(s - m^2_N)^2 + \Gamma^2_N m^2_N}\right] \; .
\label{eq: sn1}
 \end{equation}
 From this, the mean free path $l$ of a $\sigma$ particle can be calculated as
 \beq
l^{-1} = n_\nu \langle \sigma_\text{el} \rangle \;,
 \eeq
 where $\langle \sigma_\text{el} \rangle$ has been computed numerically. We will treat the motion of a dilaton particle in the supernova as a random walk process with step size $l$. Note that for the parameter range of interest (see Fig.~\ref{fig: sigma_LT}), the lifetime of $\sigma$ is much smaller than the timescale of the supernova $(\tau_\sigma \ll t_{SN} \sim 10~\text{s})$. Therefore, the number of steps in the random walk should be taken as $\gamma c\tau_\sigma/l$, resulting in
 \begin{equation}
{\bar{d}} \sim \sqrt{\gamma c \tau_\sigma l} \sim \sqrt{c \tau_\sigma l} \;.
\end{equation}
Since $m_\sigma \gg T_\mathrm{SN}$ in most of the parameter space, we drop the factor of $\gamma \sim 1 + \mathcal{O}(T_\mathrm{SN}/m_\sigma)$. Contours of $\bar{d}/L_{SN}$ are shown in Fig.~\ref{fig: sn_esc}. As can be seen, in the region where a considerable fraction of energy is transferred into the hidden sector, $\bar{d}/L_{SN} \lesssim 10^{-6}$. As a result, the $\sigma$ particles that are produced are trapped within the core. Therefore, anomalous cooling of supernovae does not place a constraint on this class of models.

 We see that in the region of interest, the hidden sector particles, once produced, are trapped inside the supernova core. Then, in the region of parameter space where $E_{\rm HS} \gtrsim E_{\rm SN}$, corresponding to the shaded region in Fig.~\ref{fig: sn_prod}, the supernova dynamics can potentially be significantly affected~\cite{Chang:2022aas}. However, given that our current understanding of supernova explosions is quite limited,  it is difficult to justify using this as the basis for a robust constraint.

\section{Conclusions}

We have studied the phenomenology of a scenario in which a strongly coupled hidden sector couples to the SM via the neutrino portal. Neutrino masses are generated via a small explicit breaking of lepton number in the hidden sector. The hidden sector is taken to be approximately conformal at short distances and has a confinement scale that lies below the electroweak scale. The lightest state in the hidden sector is the dilaton.
Since the hidden sector couples to the SM via the neutrino portal, the composite singlet neutrino $N$ is the most likely state to be produced in collider experiments. Although $N$ couples to the weak gauge bosons of the SM just like a conventional HNL, it predominantly decays to a dilaton and a neutrino. The dilaton is naturally long-lived because of angular momentum considerations, but eventually decays back to the SM. This decay may be visible or invisible, depending on the confinement scale and the $\nu$-$N$ mixing angle.

There are constraints on the parameter space of this scenario, both from beam dump and collider searches and from cosmological and astrophysical considerations. Hidden sector particles can be searched for in the rare decays of pions, kaons and charmed mesons at beam dump experiments, and B-mesons at collider experiments. Hidden sector particles can also be produced directly from the decay of weak gauge bosons at the LHC. In the near future, the most promising experiments for probing new regions in the parameter space of this class of models are FASER 2, SHiP, and Belle II. Neutrino telescopes such as IceCube are also likely have some sensitivity to this scenario~\cite{Airen:2025uhy}, but a careful analysis of their reach is left for future work.


Recently, a stochastic gravitational wave background has been observed at the nanohertz scale in multiple pulsar timing arrays~\cite{NANOGrav:2023gor,EPTA:2023fyk,Reardon:2023gzh,Xu:2023wog}. Although this signal admits an astrophysical explanation as arising from inspiralling supermassive black holes, a better fit to the data appears to be provided by first-order phase transitions~\cite{NANOGrav:2023hvm}. Strongly coupled hidden sectors that are conformal in the ultraviolet and that confine at scales in the range from 100 MeV to a few GeV have been proposed as a possible source of such a phase transition~\cite{Megias:2023kiy,Fujikura:2023lkn,Ferrante:2023bcz}. Remarkably, the scenario we have considered appears to possess all the general features necessary to explain the observed signal. However, a careful study of this possibility is left for future work.

\section*{Acknowledgements}
The research of ZC is supported by the National Science Foundation under Grant Number PHY-2514660. The research of CK and RPRS is supported by the National Science Foundation under Grant Number PHY-2210562.


\appendix
\makeatletter
\renewcommand{\@seccntformat}[1]{Appendix \csname the#1\endcsname\quad}
\makeatother
\renewcommand{\thesection}{\Alph{section}}
\numberwithin{equation}{section}


\section{Regulating the Unparticle Propagator}

 For scaling dimensions $\Delta_S \geq 2$, the left hand side of 
Eqn.~(\ref{georgi-norm}) diverges, and so this expression for the Fourier 
transform of the two-point function is no longer valid. In this 
appendix, we discuss how to regulate this expression. We follow the procedure outlined in Ref.~\cite{Ahmed:2024hpg}. It is convenient 
to define,
 \begin{equation}
\Pi_F(p^2) \equiv \int d^4x \; e^{ipx}
\langle 0|T\left[ \mathcal{O}_{S}^{\dagger}(x) \mathcal{O}_{S}(0)\right] |0 \rangle \;.
 \end{equation}
 We begin from the spectral representation of $\Pi_F(p^2)$,
 \begin{equation}
\Pi_F(p^2) =  \frac{A_{\Delta_S}}{2 \pi} 
             \int_0^\infty dM^2 (M^2)^{\Delta_S - 2}
             \frac{i}{p^2 - M^2 + i \epsilon}\,.
 \end{equation}
 For scaling dimension $\Delta_S \geq 2$, the expression on the right 
hand side is divergent. For $2 \leq \Delta_S < 3$, we may regulate this
by noting that the difference,
 \begin{equation}
 \label{georgi-norm-reg1}
\Pi_F(p^2) - \Pi_F(p_0^2) = \frac{A_{\Delta_S}}{2 \pi} 
\int_0^\infty dM^2 (M^2)^{\Delta_S - 2}
\frac{i(p_0^2 - p^2)}
{\left(p^2 - M^2 + i \epsilon\right)
 \left(p_0^2 - M^2 + i \epsilon \right)} \; ,
 \end{equation}
 is finite. Here $\Pi_F(p_0^2)$ is the value of $\Pi_F(p^2)$ at some reference 
off-shell momentum $p_0^2 < 0$. The difference can be evaluated explicitly as
\begin{equation}
\label{georgi-norm-reg2-explicit}
\Pi_F(p^2)-\Pi_F(p_0^2) = -\frac{A_{\Delta_S}}{2i\sin(\Delta_S\pi)}\left[\left(-p^2-i\epsilon\right)^{\Delta_S-2}-\left(-p_0^2-i\epsilon\right)^{\Delta_S-2}\right]\,
\end{equation}
which is regular as $\Delta_S\rightarrow 2$. We show the final result
\begin{equation}
\label{georgi-norm-reg2-limit}
\lim_{\Delta_S\rightarrow 2}\left[\Pi_F(p^2)-\Pi_F(p_0^2)\right] = -\frac{A_{\Delta_S}}{2i\pi}\left[\log\left(\frac{p^2}{|p_0^2|}\right)-i\pi\right]\,.
\end{equation} 
Note that, as expected, the absorptive part continues to agree with Eqn.~(\ref{georgi-norm-Im}) even after the two-point function has been regulated.

We can rewrite Eqn.~(\ref{georgi-norm-reg1}) as
 \begin{equation}
 \label{georgi-norm-reg2}
\Pi_F(p^2) = \Pi_F(p_0^2) + \frac{A_{\Delta_S}}{2 \pi} 
\int_0^\infty dM^2 (M^2)^{\Delta_S - 2}
\frac{i(p_0^2 - p^2)}
{\left(p^2 - M^2 + i \epsilon\right)
 \left(p_0^2 - M^2 + i \epsilon \right)} \,,
 \end{equation}
 where, in this expression, $\Pi_F(p_0^2)$ is to be treated as a free 
parameter that incorporates the unknown ultraviolet physics. Since 
$p_0^2 < 0$, the absorptive part of Eqn.~(\ref{georgi-norm-reg2}) is 
exactly the same as that of Eqn.~(\ref{georgi-norm}), and so we are 
justified in continuing to normalize the operator $\mathcal{O}_S(x)$ as 
in Eqn.~(\ref{georgi-norm-Im}).

 For $\Delta_S \geq 2$, the Higgs portal interaction in 
Eqn.~(\ref{HPortal1}) gives rise to a divergent contribution to the 
amplitude for the scattering of Higgs fields, $H H \rightarrow H H$ at 
order $\hat{g}^2$. The corresponding matrix element is of the schematic 
form,
 \begin{equation}
\label{matrix-element}
\langle p_1, p_2 | 
\int d^4 x H^{\dagger}(x) H(x) \mathcal{O}_S(x) 
\int d^4 y H^{\dagger}(y) H(y) \mathcal{O}_S(y) 
| p_A, p_B \rangle \; ,
 \end{equation}
 which scales as 
\begin{equation}
 \Pi_F([p_1 - p_A]^2) + \Pi_F([p_2 - p_A]^2) \,.
 \end{equation}
 This corresponds to a divergent contribution to the Higgs quartic 
operator in the SM, $(H^{\dagger} H)^2$. Once $\Pi_F(p^2)$ has been 
regulated by the procedure in Eqns.~(\ref{georgi-norm-reg1}) and 
(\ref{georgi-norm-reg2}), this contribution is also finite. We can 
incorporate this into a conventional renormalization scheme such as 
momentum subtraction. To illustrate this, we write the Higgs quartic 
term as the sum of a renormalized parameter and a counterterm,
 \begin{equation}
- \mathcal{L} \supset \lambda_H (H^{\dagger} H)^2 +
                      \delta_{\lambda_H} (H^{\dagger} H)^2 \; ,
 \end{equation}
 where the value of the renormalized parameter $\lambda_H$ is set by the 
scattering amplitude at the reference scale $p_0^2$. The 
counterterm $\delta \lambda_H$ is given by
 \begin{equation}
\delta_{\lambda_H} = + 2 i \frac{\hat{g}^2}{(M^2)^{\Delta_s - 2}} \Pi_F(p_0^2) + {\rm SM \; \; contributions} \; ,
 \end{equation}
 so that the sum of the (regulated) contributions to the amplitude from 
the interaction in Eqn.~(\ref{HPortal1}) and from the SM cancel exactly 
against the counterterm contribution at the scale $p_0^2$, but sum to a 
nonzero value at other scales.

 For $\Delta_S \geq 3$, the right hand side of 
Eqn.~(\ref{georgi-norm-reg2}) is still divergent and additional 
regulation is needed. For $3 \leq \Delta_S < 4$, we note that the 
difference
  \begin{equation}
 \label{georgi-norm-reg3}
\left\{\Pi_F(p^2) - \Pi_F(p_0^2)
- \Pi_F'(p_0^2)\left[p^2 - p_0^2\right]\right\} = \frac{A_{\Delta_S}}{2 \pi} 
\int_0^\infty dM^2 (M^2)^{\Delta_S - 2}
\frac{i(p_0^2 - p^2)^2}
{\left(p^2 - M^2 + i \epsilon\right)
 \left(p_0^2 - M^2 + i \epsilon \right)^2}
 \end{equation}
 is finite. This difference can also be evaluated explicitly as 
\begin{equation}
\begin{gathered}
\left\{\Pi_F(p^2) - \Pi_F(p_0^2)
- \Pi_F'(p_0^2)\left[p^2 - p_0^2\right]\right\}  = -\frac{A_{\Delta_S}}{2i\sin(\Delta_S\pi)} \\
 \times \left[\left(-p^2-i\epsilon\right)^{\Delta_S-2}-\left(-p_0^2-i\epsilon\right)^{\Delta_S-2} -(\Delta_S-2)(p_0^2-p^2)(-p_0^2-i\epsilon)^{\Delta_S-3}\right]\,,
\end{gathered}
\label{georgi-norm-reg3-explicit}
\end{equation} and it can be shown that this difference is regular as $\Delta_S\rightarrow 3$. We show the final result
\begin{equation}
\label{georgi-norm-reg3-limit}
\lim_{\Delta_S\rightarrow 3}\left\{\Pi_F(p^2) - \Pi_F(p_0^2)
- \Pi_F'(p_0^2)\left[p^2 - p_0^2\right]\right\} = -\frac{A_{\Delta_S}}{2i\pi}\left[p^2\log\left(\frac{p^2}{|p_0^2|}\right)-i\pi p^2+(p_0^2-p^2)\right]\,.
\end{equation}
Note the regulation of the two-point function has not affected the absorptive part of the two-point function, which continues to agree with Eqn.~(\ref{georgi-norm-Im}). Eqn.~(\ref{georgi-norm-reg3}) can be rewritten as an expression for 
$\Pi_F(p^2)$,
   \begin{equation}
 \label{georgi-norm-reg4}
\Pi_F(p^2) = \Pi_F(p_0^2)
+ \Pi_F'(p_0^2)\left[p^2 - p_0^2\right] + \frac{A_{\Delta_S}}{2 \pi} 
\int_0^\infty dM^2 (M^2)^{\Delta_S - 2}
\frac{i(p_0^2 - p^2)^2}
{\left(p^2 - M^2 + i \epsilon\right)
 \left(p_0^2 - M^2 + i \epsilon \right)^2} \; .
 \end{equation}
 Here $\Pi_F(p_0^2)$ and and $\Pi_F'(p_0^2)$ are to be 
treated as free parameters that incorporate the unknown 
ultraviolet physics. Once again the absorptive part of 
Eqn.~(\ref{georgi-norm}), given in Eqn.~(\ref{georgi-norm-Im}), is unaffected by the regulation procedure. This 
allows us to continue to normalize the operator $\mathcal{O}_S(x)$ as in 
Eqn.~(\ref{georgi-norm-Im}).

For $\Delta_S \geq 3$, the matrix element in Eqn.~(\ref{matrix-element}) 
gives ultraviolet divergent contributions, not just to the operator 
$(H^{\dagger} H)^2$, but also to $(H^{\dagger} H)\partial^2(H^{\dagger} 
H)$. However, once $\Pi_F(p^2)$ has been regulated following the 
procedure in Eqns.~(\ref{georgi-norm-reg3}) and (\ref{georgi-norm-reg4}), 
these contributions are also finite. In a renormalization scheme based 
on momentum subtraction, we will now require counterterms corresponding to 
both $(H^{\dagger} H)^2$ and $(H^{\dagger} H)\partial^2(H^{\dagger} H)$.

\section{Matrix Elements for Production}
\label{app: matel}

In this section, we present the helicity-averaged squared matrix elements for various processes in which a conventional HNL of mass $m_N$ is produced in the final state. We employ the notation and phenomenological parameters given in Ref.~\cite{Bondarenko:2018ptm}. In each case, the labeling of momenta $p_i$ and invariant masses $m_{ij\cdots}$ is based on the order in which the final state particles are written down.

In the case of leptonic decays of a pseudoscalar meson $\mathfrak{m}$, we have
\begin{equation}
     \overline{|\mathcal{M}|^2}(\mathfrak{m} \rightarrow \ell N) = 2G^2_F |U_{N\ell}|^2|V_{ij}|^2 f^2_\mathfrak{m} \left[m^2_\mathfrak{m} (m^2_N+m^2_\ell) - (m^2_N-m^2_\ell)^2\right]\,,
\end{equation}
where $f_\mathfrak{m}$ is the decay constant of the pseudoscalar meson, and $V_{ij}$ is the corresponding CKM matrix element.
In the case of semi-leptonic decays of pseudoscalar meson $\mathfrak{m}$ to a final state containing another pseudoscalar meson $\mathfrak{m}'$, we have
\begin{equation}
    \overline{|\mathcal{M}|^2}(\mathfrak{m} \rightarrow \mathfrak{m}' \ell N) = 4G^2_F|U_{N\ell}|^2|V_{ij}|^2 \left[2(p_2 \cdot F)(p_3 \cdot F) - (p_2 \cdot p_3)F^2\right]\,.
\end{equation}
Here $F^\mu$ denotes the hadronic matrix element corresponding to the $\mathfrak{m}(p_0) \rightarrow \mathfrak{m}'(p_1)$ transition, given by
\begin{equation}
    F^\mu(p_0, p_1) = f_+(q^2)\left[p^\mu_0 + p^\mu_1 - \frac{\Delta}{q^2}q^\mu\right] + f_0(q^2) \frac{\Delta}{q^2}q^\mu\,,
\end{equation}
where $q = p_0 - p_1 = p_2 + p_3$ and $\Delta = m^2_\mathfrak{m}-m^2_{\mathfrak{m}'}$.

\section{Decay Widths Into Unparticles}
\label{app: uN_prod}

Consider the decay of a particle $X$ of mass $M$ into a final state that contains both elementary particles and unparticles, $X \rightarrow \mathcal{U}_N Y$, where $Y$ represents the $N-1$ elementary particles in the final state. In order to compare the decay width for this process with that with a conventional HNL in the final state, $(X \rightarrow N Y)$, let us assume that the amplitude involves just a single portal interaction term, giving rise to a factor of $\hat{\lambda}v_\mathrm{EW}/m_\mathcal{U}$. As a result, the differential decay width for this process can be written as
\begin{equation}
    d\Gamma = \frac{|\hat{\lambda} v_\mathrm{EW}|^2}{M^{2\Delta_N - 3}_\mathrm{UV}}\times |\mathcal{M}'|^2 \times \frac{d\Phi_\mathcal{U}}{m^2_\mathcal{U}}  \times \frac{(2\pi)^4}{2M} d\Phi_{N-1}\,,
\end{equation}
where $\mathcal{M}'$ represents the rest of the amplitude apart from the portal coupling. The term $d\Phi_\mathcal{U}$, which arises from the unparticle phase space, takes the form
\begin{equation}
    \begin{split}
        d\Phi_\mathcal{U}   &= \mathcal{A}_{\Delta_N - \frac{1}{2}}\times \frac{d^4 p_\mathcal{U}}{(2\pi)^4} \times \Theta(p^0_\mathcal{U})                       \Theta(p^2_\mathcal{U} - \Lambda^2) \left(p^2_\mathcal{U} - \Lambda^2\right)^{\Delta_N - \frac{5}                       {2}}\\
                            &= \frac{\mathcal{A}_{\Delta_N - \frac{1}{2}}}{2\pi}\times \frac{d^3\vec{p}_\mathcal{U}}{(2\pi)^3 2E_\mathcal{U}} \times dm^2_\mathcal{U} \times\Theta (m^2_\mathcal{U} - \Lambda^2) (m^2_\mathcal{U} - \Lambda^2)^{\Delta_N - \frac{5}{2}}\,,
    \end{split}
\end{equation}
where $E_\mathcal{U} = \sqrt{\vec{p}^2_\mathcal{U} + m^2_\mathcal{U}}$. As a result, we are left with
\begin{equation}
    \begin{split}
        d\Gamma = \frac{|\hat{\lambda} v_\mathrm{EW}|^2}{M^{2\Delta_N - 3}_\mathrm{UV}} \times \frac{\mathcal{A}_{\Delta_N - \frac{1}{2}}}{2\pi}  \times \frac{dm^2_\mathcal{U}}{m^2_\mathcal{U}} (m^2_\mathcal{U} - \Lambda^2)^{\Delta_N - \frac{5}{2}} \times \frac{(2\pi)^4}{2M} |\mathcal{M}'|^2d\Phi_N\,,
    \end{split}
\end{equation}
where $d\Phi_N$ represents the N-body phase space with the mass of $\mathcal{U}_N$ fixed at $m_\mathcal{U}$. Using Eqn.~(\ref{lambdascaling}), we can write this as
\begin{equation}
\begin{split}
    d\Gamma &= \frac{\mathcal{A}_{\Delta_N - \frac{1}{2}}}{2\pi C^2_\lambda}\times \frac{dm^2_\mathcal{U}}{m^2_\mathcal{U}}\times \left(\frac{m^2_\mathcal{U}}{\Lambda^2}-1\right)^{\Delta_N - 5/2}\times \frac{(2\pi)^4}{2M} |\mathcal{M}|^2 d\Phi_N\\
    &= \frac{\mathcal{A}_{\Delta_N - \frac{1}{2}}}{2\pi C^2_\lambda}\times \frac{dm^2_\mathcal{U}}{m^2_\mathcal{U}}\times \left(\frac{m^2_\mathcal{U}}{\Lambda^2}-1\right)^{\Delta_N - 5/2}\times \widetilde{d\Gamma}(m_\mathcal{U})\,,
\end{split}
\end{equation}
where $\widetilde{d\Gamma}(m_\mathcal{U})$ is the corresponding differential decay width in the case of a conventional HNL of mass $m_\mathcal{U}$ (including the factor of $|U_{N\ell}|^2$). This can be integrated over, leading to the final result for the differential decay width into unparticles,
\begin{equation}
    d\Gamma(X \rightarrow \mathcal{U}_N Y) = \frac{\mathcal{A}_{\Delta_N - \frac{1}{2}}}{2\pi C^2_\lambda}\times \frac{dm^2_\mathcal{U}}{m^2_\mathcal{U}}\times \left(\frac{m^2_\mathcal{U}}{\Lambda^2}-1\right)^{\Delta_N - 5/2}\times \widetilde{\Gamma}(m_\mathcal{U})\,.
\label{eq: dgamma_uN}
\end{equation}

\section{Monte Carlo Integration and Sampling}
\label{app: MC}

The differential decay width for a decay to N final state particles, in which all the final state particles have definite masses, is given by
\begin{equation*}
    \widetilde{d\Gamma} = \frac{(2\pi)^4}{2M} \times |\mathcal{M}|^2 \times d\Phi_N.
\end{equation*}
This can be decomposed into sequential two-body decays to obtain
\begin{equation}
    \widetilde{d\Gamma} = \frac{(2\pi)^{3N-2}}{2M} \times |\mathcal{M}|^2 \times dm^2_{(1)} \cdots dm^2_{(N-2)} \times d\phi_{(0)} \cdots d\phi_{(N-2)} \,,
\end{equation}
where $m^2_{(i)} = \left(p_0 - \sum \limits_{k=1}^{i}p_k\right)^2$ and $d\phi_{(i)}$ is the two-body phase space corresponding to $m_{(i)} \rightarrow m_{i+1} + m_{(i+1)}$, with $m_0 = M$. Expressing the two-body phase space $d\phi_{(i)}$ in terms of the solid angles in the center of mass frame of the $m_{(i)}$, we obtain
\begin{equation}
    \widetilde{d\Gamma} = \frac{|\mathcal{M}|^2}{4M (4\pi)^{3N-4}}\times \left[\prod\limits_{i=0}^{N-2} \frac{\lambda^{1/2}_{(i)} d\Omega_{(i)}}{m^2_{(i)}}\right]\times \left[\prod\limits_{j=1}^{N-2} dm^2_{(j)}\right]\,.
\end{equation}

Adopting a sampling scheme where $\left\{\cos\theta_{(i)}, \phi_{(i)}\right\}$ are sampled uniformly and independently and the $m^2_{(j)}$ are sampled sequentially (i.e., the range from which $m^2_{(j)}$ is sampled depends on the sampled value of $m^2_{(j-1)}$), the decay width can be estimated as an average of weights $\Gamma \approx \langle \widetilde{w}\rangle $ over the sampled events, where the weight of an event $\widetilde{w_\alpha}$ is given by
\begin{equation}
    \widetilde{w_\alpha} = \frac{1}{4M (4\pi)^{2N-3}}\times \left\{|\mathcal{M}|^2 \times\left[\prod\limits_{i=0}^{N-2} \frac{\lambda^{1/2}_{(i)}}{m^2_{(i)}}\right]\times \left[\prod\limits_{j=1}^{N-2} \Delta m^2_{(j)}\right]\right\}_{\alpha}\,.
\end{equation}
We extend the same technique to the case where the final state contains an unparticle $\mathcal{U}_N$. Due to the third term on the right-hand side in Eqn.~(\ref{eq: dgamma_uN}), the distribution in $m_\mathcal{U}$ is highly biased towards $m_\mathcal{U} \rightarrow \Lambda$ for $\Delta_N < 5/2$, rendering a uniform sampling in $m_\mathcal{U}$ inefficient. Therefore, we instead uniformly sample the variable $x$ defined by the transformation,
\begin{equation}
    x = \left[\frac{m^2_\mathcal{U} - \Lambda^2}{m^2_\mathrm{max} - \Lambda^2}\right]^{\Delta_N - 3/2}\,,
\end{equation}
where $m_\mathrm{max}$ is the maximum value of $m_\mathcal{U}$ that is allowed kinematically. Then, the weight of an event is given by
\begin{equation}
    w_\alpha = \frac{\mathcal{A}_{\Delta_N - \frac{1}{2}}}{4\pi C^2_\lambda (2\Delta_N - 3)} \times \frac{\Lambda^2}{m^2_\mathcal{U}} \times \left(\frac{m^2_\mathrm{max}}{\Lambda^2} - 1\right)^{\Delta_N - \frac{3}{2}}\times \widetilde{w_\alpha}\,.
\end{equation}
We also use these weights to sample unweighted events using the acceptance-rejection method, where events are accepted with a probability of $w_\alpha/w_\mathrm{max}$. The maximum weight $w_\mathrm{max}$ is estimated during the evaluation of $\langle w \rangle$. We use the \texttt{pyhepmc} API~\cite{Verbytskyi:2020sus} for efficient computation of weights, as well as storage and retrieval of events.


\bibliography{bib_NPPS}{}
\bibliographystyle{aabib}

\end{document}